\documentclass[aps,prl,10pt,floatfix,showpacs,superscriptaddress,longbibliograph,twocolumn]{revtex4-2}
\usepackage[T1]{fontenc}
\usepackage{amsmath}
\usepackage{amssymb}
\usepackage[english]{babel}
\usepackage{booktabs}
\usepackage{graphicx}
\usepackage[colorlinks=true,urlcolor=blue,citecolor=blue,linkcolor=blue]{hyperref} 
\usepackage{algorithm}  
\usepackage{algpseudocode}
\usepackage{multirow}
\usepackage{newtxmath}
\usepackage{newtxtext}
\usepackage{subfigure}
\usepackage{arydshln}
\usepackage[dvipsnames]{xcolor}
\usepackage{siunitx}

\newcommand{\s}{\mathbf{s}}
\newcommand{\x}{\mathbf{x}}

\newcommand{\btheta}{\boldsymbol{\theta}}

\newcommand{\KLD}[2]{D_{\text{KL}} \left(#1 \parallel #2 \right) }

\begin{document}
\title{Tensor networks for unsupervised machine learning}

\author{Jing Liu}
\affiliation{School of Systems Science, Beijing Normal University, Beijing 100875, China}

\author{Sujie Li}
\affiliation{CAS Key Laboratory for Theoretical Physics, Institute of Theoretical Physics, Chinese Academy of Sciences, Beijing 100190, China}
\affiliation{School of Physical Sciences, University of Chinese Academy of Sciences, Beijing 100049, China}

\author{Jiang Zhang}
\affiliation{School of Systems Science, Beijing Normal University, Beijing 100875, China}
\affiliation{Swarma Research, Beijing 102308, China}

\author{Pan Zhang}
\email{panzhang@itp.ac.cn}
\affiliation{CAS Key Laboratory for Theoretical Physics, Institute of Theoretical Physics, Chinese Academy of Sciences, Beijing 100190, China}
\affiliation{School of Fundamental Physics and Mathematical Sciences, Hangzhou Institute for Advanced Study, UCAS, Hangzhou 310024, China}
\affiliation{International Centre for Theoretical Physics Asia-Pacific, Beijing/Hangzhou, China}

\begin{abstract}
Modeling the joint distribution of high-dimensional data is a central task in unsupervised machine learning.
In recent years, many interests have been attracted to developing learning models based on tensor networks, which have the advantages of a principle understanding of the expressive power using entanglement properties, and as a bridge connecting classical computation and quantum computation.
Despite the great potential, however, existing tensor network models for unsupervised machine learning only work as a proof of principle, as their performance is much worse than the standard models such as restricted Boltzmann machines and neural networks.
In this Letter, we present autoregressive matrix product states (AMPS), a tensor network model combining matrix product states from quantum many-body physics and autoregressive modeling from machine learning.
Our model enjoys the exact calculation of normalized probability and unbiased sampling.
We demonstrate the performance of our model using two applications, generative modeling on synthetic and real-world data, and reinforcement learning in statistical physics.
Using extensive numerical experiments, we show that the proposed model significantly outperforms the existing tensor network models and the restricted Boltzmann machines, and is competitive with state-of-the-art neural network models.
\end{abstract}

\maketitle

\paragraph{Introduction.}
Unsupervised learning aims to understand the high-dimensional data $\x$ with rich structures, by learning a probability distribution $P(\x)$ from the data.
The central challenge is how to represent the joint distribution with an exponential number of entries using a polynomial number of parameters and how to efficiently sample from it.
Indeed, even computing the normalization constant for a general joint distribution belongs to the \#P-hard problem, and has attracted enormous efforts~\cite{bishop2006pattern,Goodfellow2016Deep} in the past decades.
From the point of view of computations, unsupervised learning shares the same challenges with statistical physics and quantum many-body physics, where the target distribution or the target quantum state one investigates lives in a huge space and is difficult to control.

In the history of machine learning, many classical unsupervised learning models were inspired by physics.
One of the most important models, the celebrated Boltzmann machine~\cite{hinton1984boltzmann,smolensky1986information}, was inspired by statistical physics, where $P(\x)$ is expressed using the Boltzmann distribution with parameters encoded in the energy function $E(\x)$,
\begin{equation}
    P(\x)=\frac{1}{Z}e^{-\beta E(\x)}.
\end{equation}
To efficiently sample from the Boltzmann distribution, the model is usually defined on a bipartite graph where efficient block-update Monte Carlo methods exist.
This is known as the restricted Boltzmann machine (RBM)~\cite{hinton2002training,hinton2006reducing} and has found numerous applications in machine learning and physics.
Another physics-inspired model is the Born machine~\cite{han2018unsupervised,cheng2018information}, where the joint distribution of variables is modeled by the squared norm of a quantum state $\Psi(\x)$, with
\begin{equation}
    P(\x)=\frac{1}{Z}\left |\Psi(\x)\right |^2,
\end{equation}
and the quantum state is represented by tensor networks~\cite{han2018unsupervised,cheng2018information,cheng2019tree,glasser2019expressive,Stoudenmire2016a,stoudenmire2018learning,cheng2021supervised,meng2020residual,liu2019machine}.
Given the probabilistic interpretations of quantum mechanics, as well as the close relations between tensor networks and quantum circuits, Born machines have potential applications in quantum machine learning.

Theoretically, tensor network models based on matrix product states (MPSs)~\cite{han2018unsupervised,perez2006matrix} and tree tensor networks (TTNs)~\cite{cheng2019tree,2006Classical} enjoy tractable normalization and a good theoretical interpretation of the representation power.
However, when applied to practical unsupervised learning tasks on the standard data set, the performance of tensor network models is much worse than the neural network models~\cite{han2018unsupervised,cheng2019tree}, even has a significant gap to the RBM which was invented more than $30$ years ago~\cite{hinton2002training,hinton2006reducing}.

In this Letter, we aim to fill this gap.
Based on the MPS, we propose an unsupervised model which maintains the advantages of tractable normalization and exact sampling, while having much stronger representation power than the MPS Born machine~\cite{han2018unsupervised}.
In the proposed model, instead of representing a joint distribution $P(\x)$ directly using a tensor network, we combine it with the autoregressive construction of joint probabilities, by factorizing the joint probability as a product of conditional probabilities.
We refer to our model as \textit{autoregressive matrix product states} (AMPS).
For the theoretical representation power, the model has a two-dimensional tensor network representation, which has similar entanglement structures to the two-dimensional tensor network representation of the RBM~\cite{li2021boltzmann}, while it has the ability to compute the exact normalized probability and unbiased direct sampling.
For applications, by varying the parameters of the conditional probabilities, i.e., tensor elements of MPSs, we can train the parametrized model distribution $P(\x)$ to accomplish various tasks.
In the generative learning tasks~\cite{Goodfellow2016Deep}, we learn the empirical distribution of the training data set in a maximum likelihood estimation procedure for density estimation and generating new samples from it.
In the reinforcement learning of statistical physics~\cite{wu2019solving}, we learn the Boltzmann distribution of a given instance with the REINFORCE algorithm~\cite{williams1992simple} to estimate the free energy and the ensemble average of an observable such as magnetization.
Using numerical experiments, we show that in these two applications, the performance of AMPS is much stronger than existing tensor network models and is competitive with the neural network models.

\begin{figure}[!htbp]
    \centering
    \includegraphics[width=\linewidth]{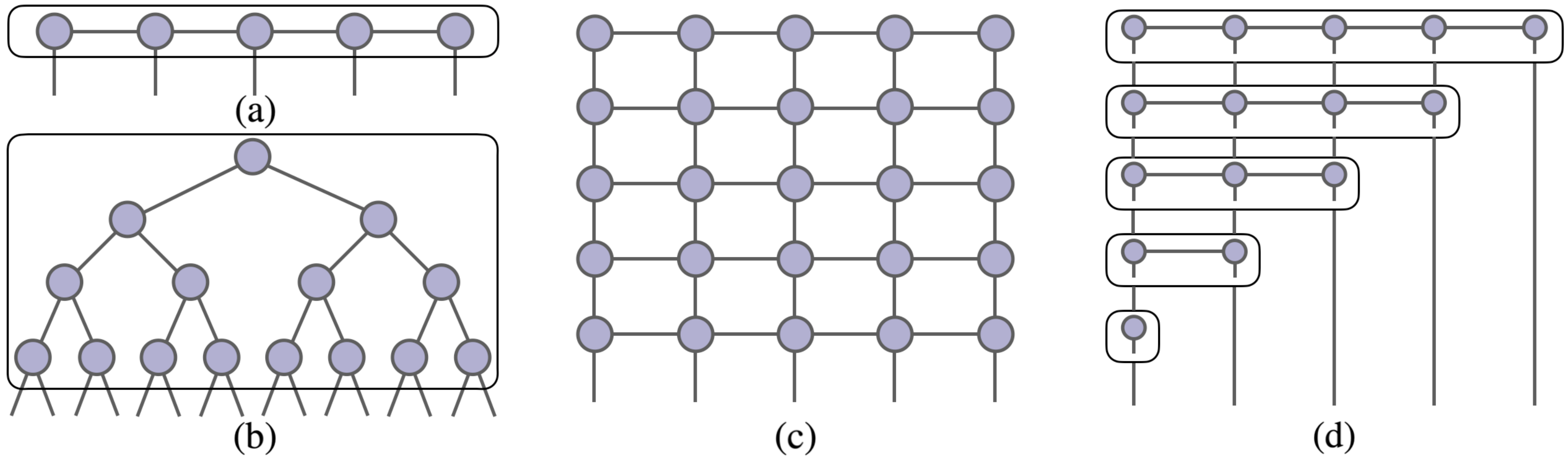}
    \caption{Schematic representation of generative models based on (a) MPS, (b) TTN, (c) RBM with the two-dimensional tensor network representation~\cite{li2021boltzmann}, and (d) AMPS. In (d), the joint probability is factorized as the product of conditional probabilities and each of them is parametrized using an MPS of a different length. In (a), (b), and (d), the solid box denotes the non-linear or linear function to normalize the tensor network.}
    \label{fig:mps}
\end{figure}

\paragraph{Autoregressive matrix product states.}
The standard tensor networks in quantum many-body physics such as MPS and TTN illustrated in Figs.~\ref{fig:mps}(a) and \ref{fig:mps}(b) have been generalized to unsupervised machine learning~\cite{han2018unsupervised,cheng2019tree,glasser2019expressive}.
The obtained generative models have the advantages of exact computation of normalized probabilities and direct sampling.
However, when applied to generative learning tasks on the standard data sets, their performances have a significant gap~\cite{han2018unsupervised,cheng2019tree} compared to the RBM and neural network models. The starting point of this work is to understand the reason behind the gap and to investigate how to fill it.

For simplicity, consider an RBM of $n$ visible variables and $m$ hidden variables without bias parameters. The hidden variables of the RBM can be explicitly marginalized out as 
\begin{equation}
    P(\x)=\frac{1}{\widetilde Z} \prod_{a=1}^m \Psi^{(a)}_{x_1 x_2 \cdots x_n},
    \label{eq:product-of-mps}
\end{equation}
where $\Psi^{(a)}_{x_1 x_2 \cdots x_n}=2\cosh(\sum_{i=1}^n W_{ia}x_i)=\prod_{i=1}^n \exp(W_{ia}x_i)+\prod_{i=1}^n \exp(-W_{ia}x_i)$ is a sum over two rank-one tensors, hence it is in a canonical polyadic (CP) form where CP rank is equal to 2.
Since a CP tensor can be converted exactly to an MPS with a bond dimension equal to its CP rank~\cite{MAL-059}, we say that an RBM can be regarded as a product of MPSs~\cite{glasser2019expressive,clark2018unifying}.
Notice that by converting each factor $\Psi^{(a)}_{x_1 x_2\cdots x_n}$ into an MPS with bond dimension 2, the total number of parameters becomes $m\times n \times 2 \times 2^2$ compared to $m\times n$ of the RBM.
However, converting the same RBM into a single MPS may require an exponentially large bond dimension.
This is straightforward from Fig.~\ref{fig:mps}(c) that if we contract the two-dimensional tensor network representation of RBM~\cite{li2021boltzmann} from top to bottom, we will eventually obtain an MPS with bond dimension $2^m$ and the total number of parameters of this single MPS has become $n\times 2 \times (2^m)^2$.
The above discussion implies that the product of MPSs exhibits higher parameter efficiency compared to a single MPS transformed from the same RBM.
However, it comes with the cost of an intractable (unbiased) sampling and computation of the normalization factor.

The autoregressive matrix product states (AMPS) model proposed in this Letter can readily take advantage of the product of MPSs but at the same time preserve the tractability of sampling and the normalization factor.
As illustrated in Fig.~\ref{fig:mps}(d), the proposed tensor network is also designed as a product of factors, with each factor $\Psi^{(i)}_{x_1 x_2 \cdots x_i}$ having a different length,
\begin{equation}
    P(\x)=\frac{1}{\widetilde Z}\Psi^{(1)}_{x_1}\Psi^{(2)}_{x_1 x_2}\cdots\Psi^{(n)}_{x_1 x_2 \cdots x_n}.
    \label{eq:auto}
\end{equation}
The difference between $\Psi^{(i)}_{x_1 x_2 \cdots x_i}$ in Eq.~(\ref{eq:auto}) and $\Psi^{(a)}_{x_1 x_2 \cdots x_n}$ in Eq.~(\ref{eq:product-of-mps}) is that $\Psi^{(i)}_{x_1 x_2 \cdots x_i}$ is designed to be normalized such that $\sum_{x_i}\Psi^{(i)}_{x_1 x_2 \cdots x_i}=1$, meaning that $\Psi^{(i)}_{x_1 x_2 \cdots x_i}$ expresses the conditional probability $P(x_i|x_1,\cdots,x_{i-1})$.
This is also known as autoregressive modeling in machine learning and it enjoys tractable normalization since the normalization factor $\widetilde{Z}$ is $1$.
We refer to the Supplemental Material~\cite{supplemental_material}\nocite{MAL-059,oseledets2011tensor,glasser2018neural,kingma2014adam,muller2007classification,Dua:2019,download,nicoli2020asymptotically} for details on the AMPS model, including the two-dimensional tensor network representation in Figs.~\ref{fig:mps}(c) and \ref{fig:mps}(d).

We propose to represent each factor using an MPS:
\begin{equation}
    \psi_{x_1 x_2 \cdots x_n} = 
    \sum_{\{\alpha_i\}}
    A^{(1)}_{x_1\alpha_0 \alpha_1}
    A^{(2)}_{x_2\alpha_1 \alpha_2} \cdots
    A^{(n)}_{x_{n}\alpha_{n-1} \alpha_n},
\end{equation}
where $A^{(i)}_{x_i \alpha_{i-1} \alpha_i}\in\mathbb R^{d\times D_{i-1}\times D_{i}}$ is a rank-3 tensor and $D_0=D_n=1$ in this case.
The first index $x_i$ corresponds to the physical variable $x_i$ with dimension $d$, and the second and third indices $\alpha_{i-1}$ and $\alpha_i$ correspond to the interaction to the previous and the next rank-3 tensor, with bond dimension $D_{i-1}$ and $D_{i}$, respectively.
In order to express a conditional distribution, $\Psi^{(i)}_{x_1 x_2 \cdots x_i}$ must be non-negative.
There are several ways to ensure this condition. One way is to normalize it by the squared norm of $\psi$, i.e., $\Psi^{(n)}_{x_1 x_2 \cdots x_n} = \mid\psi_{x_1 x_2 \cdots x_n}\mid^2 / {\widehat Z}$, where $\widehat Z$ is the normalizing factor (same as the MPS Born machines~\cite{han2018unsupervised}).
Another way is to use an exponential parametrization of $\psi$: $\Psi^{(n)}_{x_1 x_2 \cdots x_n} = e^{\psi_{x_1 x_2\cdots x_n}}/\sum_{x_n} e^{\psi_{x_1 x_2 \cdots x_n}}$.
Moreover, one can impose a non-negative condition for every tensor element in $\psi$.
In Fig.~\ref{fig:mps}(d) we use a solid box to represent the linear or non-linear function that normalizes the tensor network.
We found that the second choice, exponential parametrization works best in practice.

\paragraph{AMPS for generative modeling.}

\begin{figure*}[!htbp]
    \centering
    \includegraphics[width=0.8\linewidth]{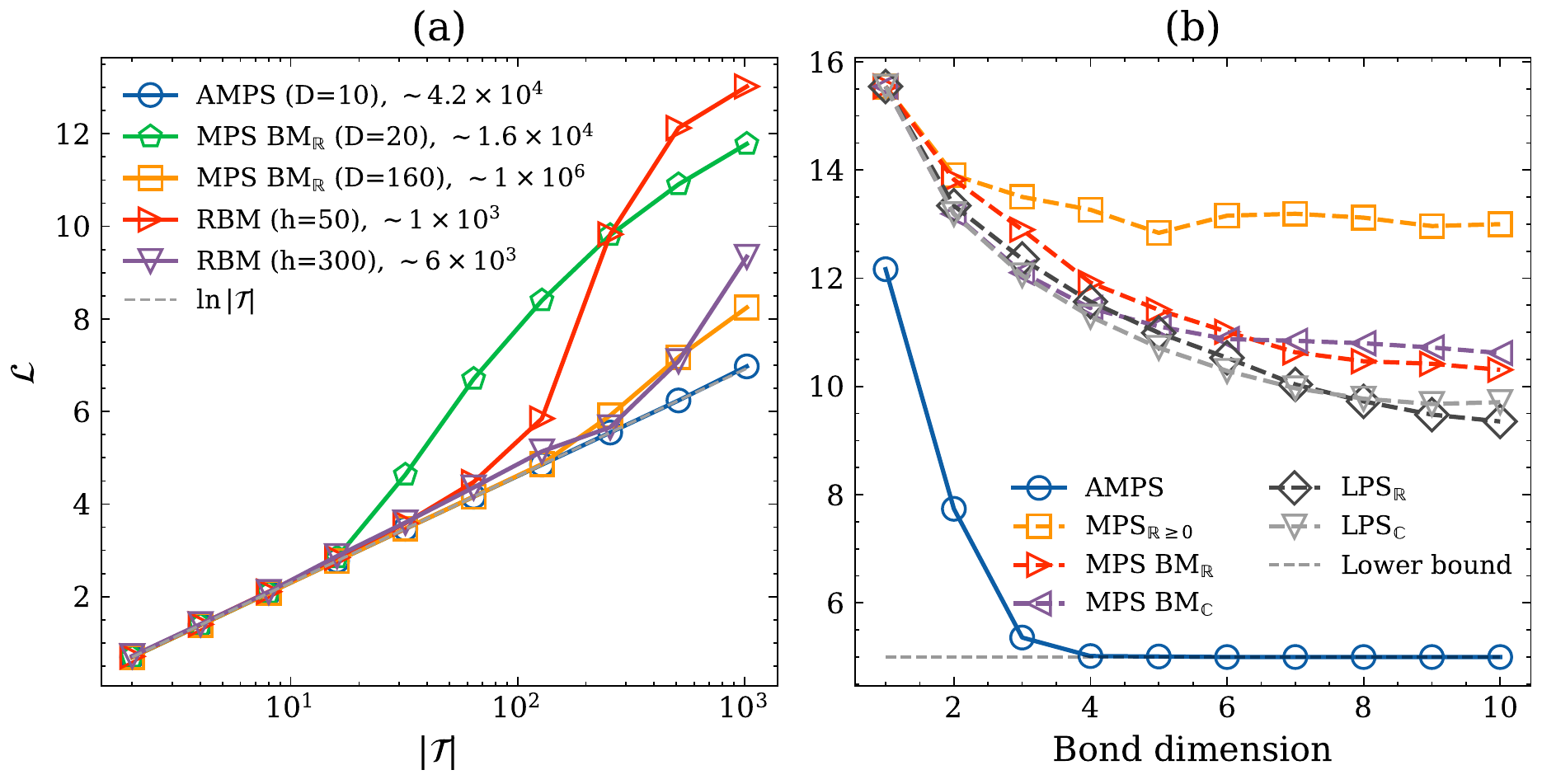}
    \caption{(a) For random datasets with $20$ variables, we plot the NLL vs the number of samples $\vert \mathcal{T}\vert$ for AMPS, MPS Born machine with real entries ($\textrm{MPS BM}_{\mathbb R}$), and RBM. We also give the number of parameters of each model in the legend. (b) For the real-world data set called Lymphography~\cite{lymphodataset}, we plot the NLL as a function of bond dimension $D$ for AMPS, Positive MPS ($\textrm{MPS}_{\mathbb R\geq 0}$), MPS Born machines with real entries ($\textrm{MPS BM}_{\mathbb R}$) and complex entries ($\textrm{MPS BM}_{\mathbb C}$), and locally purified states with real entries ($\textrm{LPS}_{\mathbb R}$) and complex entries ($\textrm{LPS}_{\mathbb C}$). The dashed line indicates the lower bound and is determined by the entropy of the empirical data distribution.}
    \label{fig:random}
\end{figure*}

Generative modeling asks to learn the parameters of $P(\x)$ as close as possible to the empirical distribution of a given data set $\mathcal{T}$ with $\vert \mathcal{T}\vert$ data samples by minimizing the negative log likelihood (NLL).
It is equivalent to minimizing the Kullback-Leibler (KL) divergence between the empirical data distribution and $P(\x)$~\cite{Goodfellow2016Deep,lecun2015deep}. 
In this work, we assume that the data variables are binary, i.e., $\x\in\{0,1\}^n$ for each data sample containing $n$ variables. 
After learning, the achieved NLL of the model quantifies the performance of the learned model in representing the training data set.
The lower bound of the NLL is the entropy of the empirical data distribution, which equals $\ln |\mathcal{T}|$ if there are no duplications.

First, we study the case when the training data are random binary samples.
How many random patterns can be remembered by the model indicates the expressive power of the model, also known as model capacity~\cite{hopfield1982neural}.
When $|\mathcal T|$ random patterns are exactly remembered, the NLL of the model would be $\ln |\mathcal T|$, and the learned distribution faithfully represents the empirical distribution with $|\mathcal T|$ nonzero values.
In the experimental evaluations, we plot the NLL obtained in AMPS as a function of the number of data samples $|\mathcal T|$ and compared it to MPS Born machines~\cite{han2018unsupervised} with the different bond dimensions and RBMs with different numbers of hidden variables, trained on the same data.
We chose $n=20$ so that the partition function of RBM, the gradients of parameters, as well as its NLL, can be computed exactly.
The results are reported in Fig.~\ref{fig:random}(a) where the lower bound is plotted in a dashed line.
We see that with more hidden variables, RBM will have more parameters ($n\times h$) and thus is more expressive, giving a smaller NLL.
Similarly, with a larger bond dimension, the MPS Born machine also contains more parameters ($nD^2$) and also obtains a smaller NLL.
In particular, with bond dimension $D\geq |\mathcal T|$, the MPS Born machine is able to faithfully clone the empirical data distribution~\cite{han2018unsupervised} and gives an NLL of $\ln|\mathcal T|$.
The figure demonstrates that with a small bond dimension $D=10$, the proposed AMPS already significantly outperforms RBM with $h=300$ hidden variables, and a Born machine with bond dimension $D=160$, exhibiting a much stronger expressive power than both the RBM and MPS Born machine.

We also test the expressive power on real-world datasets studied in Ref.~\cite{glasser2019expressive} and evaluated its performance with other tensor network models such as the MPS Born machine with complex entries and the locally purified states (LPS)~\cite{glasser2019expressive}.
The data variables in real-world datasets are categorical, i.e., $\x \in \{0, 1, \cdots, d-1\}^n$ where $d>2$ denotes the number of categories.
The results on the Lymphograph data set~\cite{lymphodataset} are shown in Fig.~\ref{fig:random}(b), where we can see that with the same bond dimension, AMPS is significantly superior to other tensor network models, obtaining much lower NLL.
In particular, with a bond dimension larger than 3, AMPS has achieved an NLL equal to the lower bound, which is already beyond the reach of other models even with a larger number of parameters.
A brief introduction to the compared tensor network models and more results are provided in the Supplemental Material~\cite{supplemental_material}, further confirming the superiority of AMPS in both performance and parameter efficiency.

\begin{table}[!htbp]
    \centering
    \caption{Negative log likelihood on the test data set (Test NLL, the smaller the better) given by AMPS on the binarized MNIST data set, compared with different tensor network models and neural network models. For RBM, $*$ stands for an approximated NLL. The total number of parameters for each model is also summarized. For the MPS Born machine and the TTN Born machine, the number of parameters is evaluated approximately using the maximum bond dimension.}
    \begin{tabular}{@{}lrr@{}}
    \toprule
    \textbf{Nonconvolution Models}  & \bf{\,\,\,\,\,\,\,\,\,Parameters} & \bf{\,\,\,\,\,\,\,\,\,Test NLL} \\
    \midrule
    MPS Born machine & $\sim$\num{15680}\text{k} & 101.5~\cite{cheng2019tree}   \\
    TTN Born machine (1d)   & $\sim$\num{127855}\text{k}  & 96.9~\cite{cheng2019tree}     \\
    TTN Born machine (2d)   & $\sim$\num{127855}\text{k}   & 94.3~\cite{cheng2019tree}     \\
    RBM      & $\sim$\num{393}\text{k} & 86.3$^*$~\cite{salakhutdinov2008quantitative}\\
    NADE    & $\sim$\num{785}\text{k}         & 88.3~\cite{uria2014deep}     \\
    MADE    & $\sim$\num{76561}\text{k}         & 86.6~\cite{germain2015made}     \\
    {AMPS}  & $\sim$\num{15680}\text{k} & {\textbf{84.1}}     \\ 
    \midrule
    \bf{Convolution Models}  &   \bf{Parameters}      & \bf{Test NLL} \\ 
    \midrule
    PixelCNN   & $\sim$\num{21}\text{k}      & 81.3~\cite{van2016pixel}     \\ 
    {Deep-AMPS}     & $\sim$\num{602}\text{k}       & {81.8}     \\ \bottomrule
    \end{tabular}
    \label{tab:nll}
\end{table}

Next, to demonstrate the generalization ability, we perform experiments on the binarized MNIST data set~\cite{binarizedMNIST}, a standard benchmark for generative models in machine learning.
The binary MNIST data set contains a total of \num{50000} training, \num{10000} testing, and \num{10000} validation images of handwritten digits, each of which is composed of $28\times 28$ binary pixels. 
We force all MPSs in the AMPS to totally share the parameters, which can greatly improve the generalization power and reduces the number of parameters (see Supplemental Material~\cite{supplemental_material} for details).
We use all \num{50000} images for training the model and the generalization power is characterized by the averaged NLL on the unobserved test data set.
The results are summarized in Table~\ref{tab:nll}, benchmarked with the MPS Born machine~\cite{han2018unsupervised}, the TTN Born machine~\cite{cheng2019tree}, and generative models that parametrize conditional probabilities using deep neural networks: the neural autoregressive distribution estimator (NADE)~\cite{larochelle2011neural}, masked autoencoder distribution estimator (MADE)~\cite{germain2015made}, and pixel convolutional neural networks (PixelCNN)~\cite{van2016pixel}.
NADE, MADE, and PixelCNN are autoregressive models with conditional probabilities represented using neural networks.
Among them, NADE and MADE are nonconvolutional models, while the PixelCNN uses convolutional neural networks and is considered one of the state-of-the-art models on the binarized MNIST data set.
The test NLL given by AMPS is 84.1 with a corresponding bond dimension $D=100$, which greatly improves over the existing tensor network methods.
Remarkably, AMPS is competitive with the generative models using deep neural networks and even outperforms the non-convolutional MADE and NADE.
We also see from Table~\ref{tab:nll} that there is an NLL gap between AMPS and PixelCNN.
The reason for the gap is clearly the absence of a convolutional structure in AMPS on the image data.
Therefore, we incorporate the convolutional structure into AMPS, which is referred to as Deep-AMPS.
The obtained NLL is very close to the result of PixelCNN.
The details of the Deep-AMPS are provided in the Supplemental Material~\cite{supplemental_material}.

\begin{figure}[!htbp]
    \centering
    \subfigure[]{
    \includegraphics[width=0.29\linewidth]{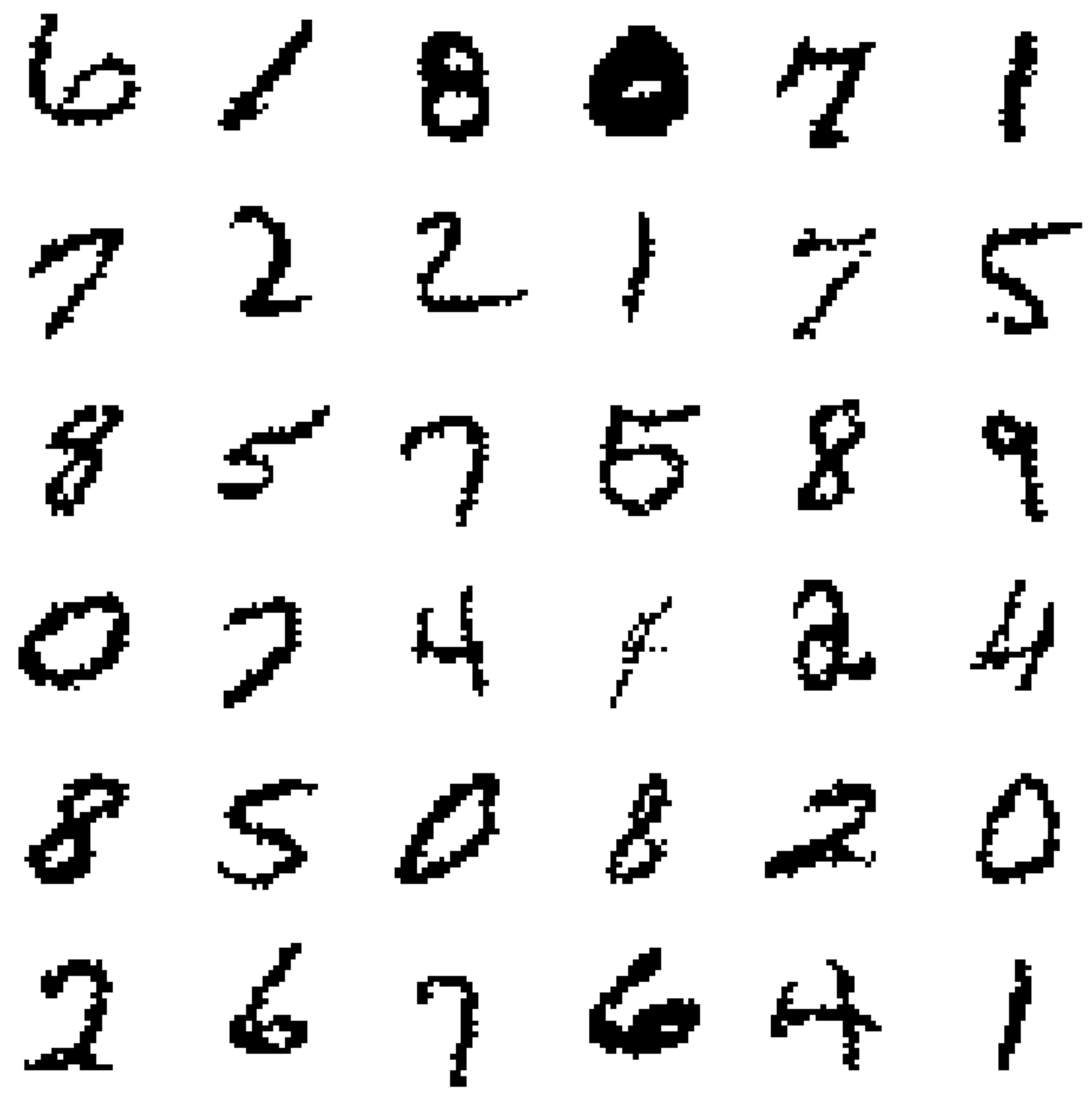}
    \label{fig:mnist}}
    \,\,
    \subfigure[]{
    \includegraphics[width=0.29\linewidth]{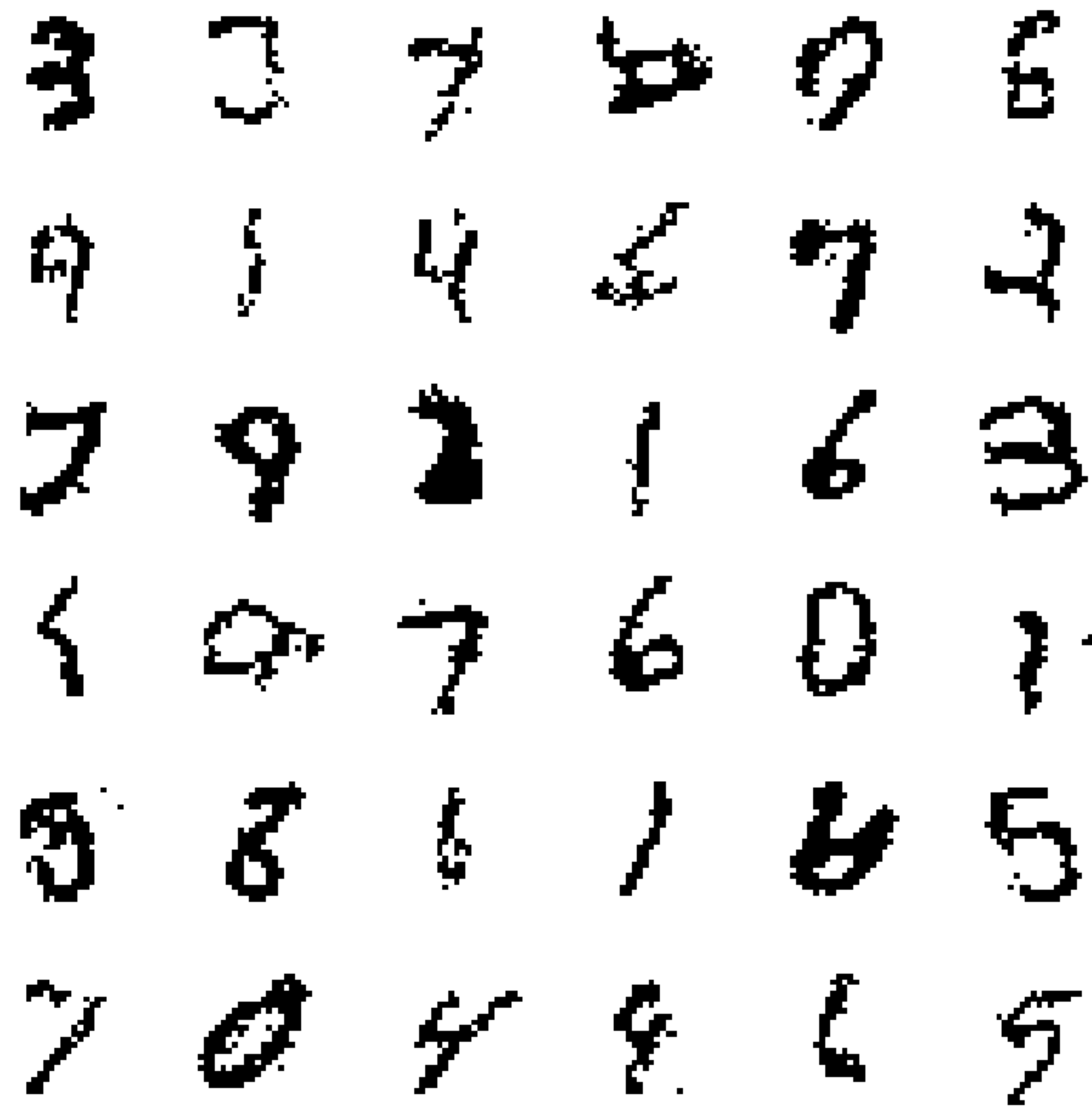}
    \label{fig:amps}}
    \,\,
    \subfigure[]{
    \includegraphics[width=0.29\linewidth]{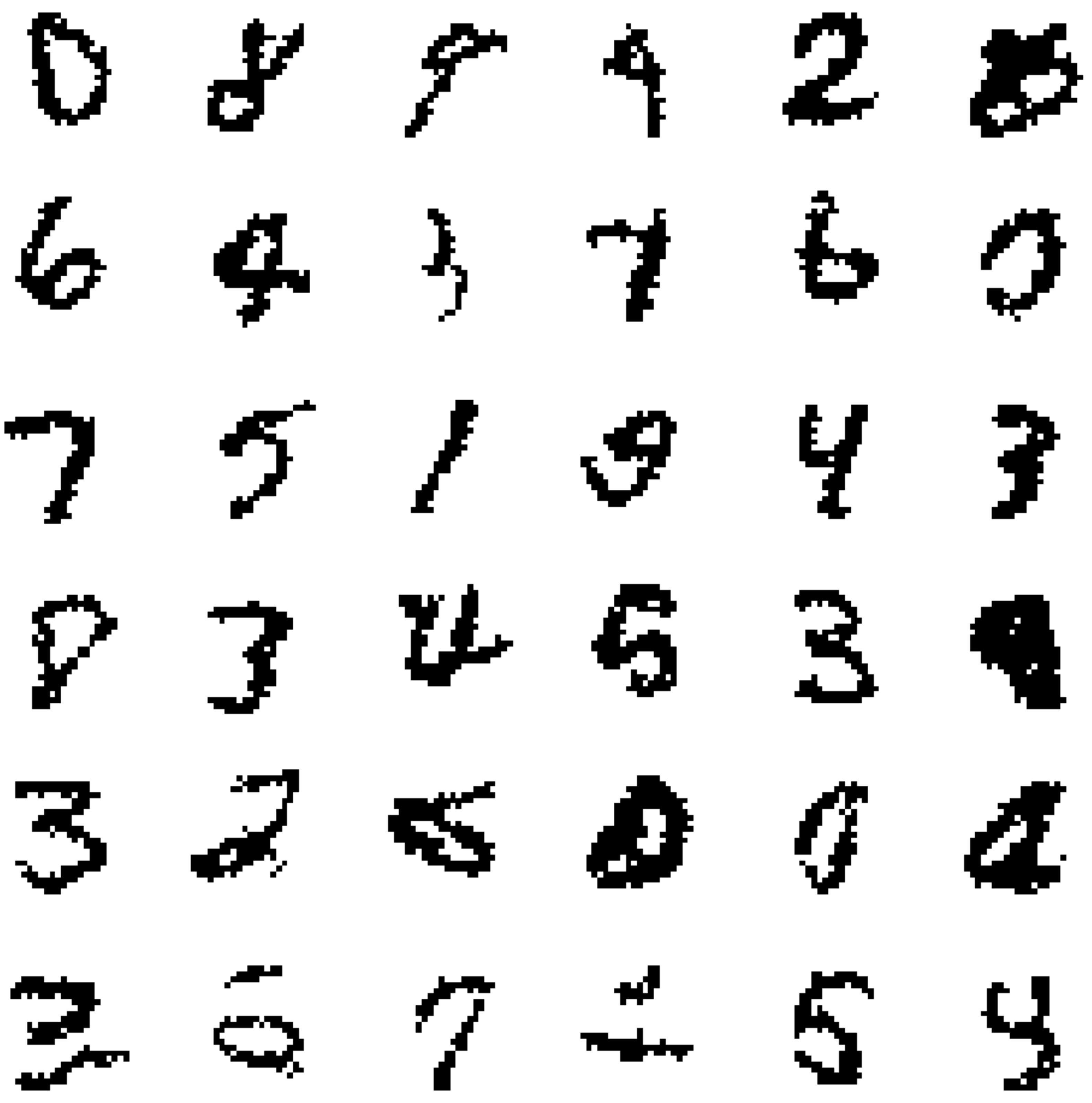}
    \label{fig:con_mps}}
    \caption{(a) Handwritten digits in the binarized MNIST data set. (b) Digits sampled from the learned AMPS ($D=100)$, $\text{test NLL}=84.1$. (c) Digits sampled from the learned Deep-AMPS (8 layers and $D=14$), test NLL$=81.8$.}
    \label{fig:samples}
\end{figure}

Owing to modeling every conditional probability in AMPS, we are able to sample pixels one by one from the learned distribution in an unbiased way, which is known as ancestral sampling~\cite{bishop2006pattern} in machine learning.
To visually illustrate the quality of images generated by the learned AMPS and Deep-AMPS, in Figs.~\ref{fig:amps} and \ref{fig:con_mps} we randomly sample 36 images from AMPS ($D=100$) and Deep-AMPS (8 layers and $D=14$), respectively.
We can see that most of the generated digits are recognizable and are visually very close to the original digits in the binarized MNIST data set in Fig.~\ref{fig:mnist}.

\paragraph{AMPS for reinforcement learning in statistical physics.}
Reinforcement learning~\cite{sutton2018reinforcement} has many applications in artificial intelligence~\cite{silver2017mastering}, as well as in physics, e.g., as a variational framework in statistical mechanics~\cite{wu2019solving,pan2019solving} and in quantum many-body physics~\cite{carleo2017solving,sharir2020deep}.
For statistical mechanics problems, reinforcement learning is used to minimize the variational free energy $F_q$ of a variational distribution $q_\theta(\s)$, an upper bound of the true free energy of the Boltzmann distribution,
\begin{equation}
    \label{eq:variational_f}
    F_q=\frac{1}{\beta}\sum_{\s} q_{\theta}(\s)\left[\beta E(\s)+\ln q_{\theta}(\s)\right],
\end{equation}
where $\s \in \{-1,+1\}^n$ is the spin configuration.
In conventional mean-field methods such as the naive mean field (NMF), the variational distribution reads $q_\theta(\s)=\prod_i q_i(s_i)$ so the variational free energy and its derivatives with respect to the parameters can be expressed analytically.
In the recently proposed variational autoregressive networks (VANs)~\cite{wu2019solving,pan2019solving}, $q_\theta(\s)$ is parametrized using deep neural networks.
In this work, we show that AMPS can also be applied as a variational distribution $q_{\textrm{AMPS}}(\s)$ for minimizing the free energy $F_q$ in Eq.~(\ref{eq:variational_f}) by reinforcement learning and achieving competitive performance compared to the neural network models.

\begin{figure}[!htbp]
    \centering
    \includegraphics[width=\linewidth]{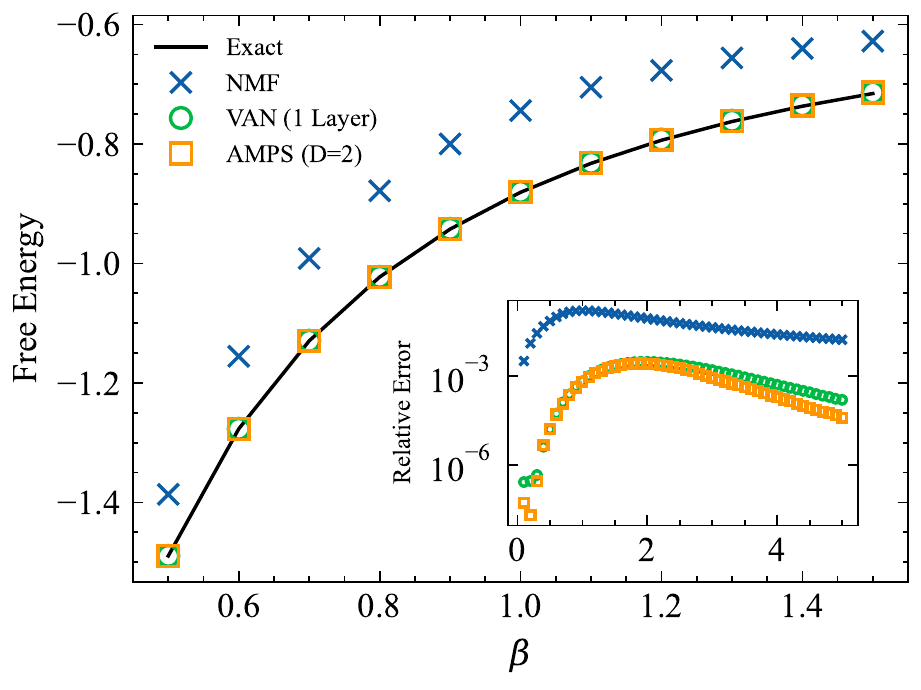}
    \caption{Free energy given by the naive mean field (NMF)~\cite{anderson1987mean}, variational autoregressive network (VAN)~\cite{wu2019solving}, and the AMPS on the SK spin-glass model~\cite{sherrington1975solvable} with 20 spins. The inset plot shows the relative error on a logarithmic scale. In the experiment, the VAN (1 layer) and the AMPS ($D=2$) have roughly the same number of parameters.}
    \label{fig:sk}
\end{figure}

To evaluate the performance, we perform experiments on the celebrated Sherrington-Kirkpatrick (SK) spin-glass model~\cite{sherrington1975solvable}.
The energy of a configuration $\s$ is defined as $E(\s) = -\sum_{1\leq i < j \leq n} J_{ij} s_i s_j$ where $n$ denotes the number of spins and the couplings $J_{ij} =J_{ji}$ follow the Gaussian distribution with $0$ mean and $1/\sqrt{n}$ variance.
We follow the experimental settings in Ref.~\cite{wu2019solving} and choose $n=20$ in order to enumerate all the $2^n$ configurations for computing the exact free energy for the evaluations.
The results are shown in Fig.~(\ref{fig:sk}). 
We can see that the AMPS with a corresponding bond dimension $D=2$ not only significantly outperforms the standard mean-field method but also achieves competitive results compared to VAN, which is considered the state-of-the-art variational free energy minimization of the SK spin-glass model. Remarkably, at both high temperatures and low temperatures, the AMPS works even better than VAN.

\paragraph{Discussions.}
We have presented the AMPS, a tensor network model for unsupervised machine learning.
The model incorporates autoregressive modeling in machine learning and matrix product states in physics.
It enjoys the exact computation of joint probability inherited from the autoregressive modeling.
Theoretically, the expressive power of the AMPS is much stronger than existing tensor network models.
Empirically, using extensive experiments we have demonstrated that the AMPS outperforms existing tensor network models in generative modeling of images and reinforcement learning in variational statistical mechanics of spin glasses.
Remarkably, its performance is competitive with the state-of-the-art deep neural network models, even better in some cases. This uncovers the great potential of tensor network methods in machine learning.
Finally, we note that with natural relations between tensor networks and quantum circuits~\cite{boixo2018characterizing,pan2020contracting,pan2022simulation}, it would be interesting to further investigate along this line for combining autoregressive modeling and quantum machine learning.

\begin{acknowledgments}
P.Z. is supported by WIUCASICTP2022, and Projects No. 11747601 and No. 11975294 of NSFC.
We provide PyTorch implementations of the algorithm in Ref.~\cite{code}.
\end{acknowledgments}

\bibliography{main.bib}

\begin{thebibliography}{50}%
\makeatletter
\providecommand \@ifxundefined [1]{%
 \@ifx{#1\undefined}
}%
\providecommand \@ifnum [1]{%
 \ifnum #1\expandafter \@firstoftwo
 \else \expandafter \@secondoftwo
 \fi
}%
\providecommand \@ifx [1]{%
 \ifx #1\expandafter \@firstoftwo
 \else \expandafter \@secondoftwo
 \fi
}%
\providecommand \natexlab [1]{#1}%
\providecommand \enquote  [1]{``#1''}%
\providecommand \bibnamefont  [1]{#1}%
\providecommand \bibfnamefont [1]{#1}%
\providecommand \citenamefont [1]{#1}%
\providecommand \href@noop [0]{\@secondoftwo}%
\providecommand \href [0]{\begingroup \@sanitize@url \@href}%
\providecommand \@href[1]{\@@startlink{#1}\@@href}%
\providecommand \@@href[1]{\endgroup#1\@@endlink}%
\providecommand \@sanitize@url [0]{\catcode `\\12\catcode `\$12\catcode
  `\&12\catcode `\#12\catcode `\^12\catcode `\_12\catcode `\%12\relax}%
\providecommand \@@startlink[1]{}%
\providecommand \@@endlink[0]{}%
\providecommand \url  [0]{\begingroup\@sanitize@url \@url }%
\providecommand \@url [1]{\endgroup\@href {#1}{\urlprefix }}%
\providecommand \urlprefix  [0]{URL }%
\providecommand \Eprint [0]{\href }%
\providecommand \doibase [0]{https://doi.org/}%
\providecommand \selectlanguage [0]{\@gobble}%
\providecommand \bibinfo  [0]{\@secondoftwo}%
\providecommand \bibfield  [0]{\@secondoftwo}%
\providecommand \translation [1]{[#1]}%
\providecommand \BibitemOpen [0]{}%
\providecommand \bibitemStop [0]{}%
\providecommand \bibitemNoStop [0]{.\EOS\space}%
\providecommand \EOS [0]{\spacefactor3000\relax}%
\providecommand \BibitemShut  [1]{\csname bibitem#1\endcsname}%
\let\auto@bib@innerbib\@empty
\bibitem [{\citenamefont {Bishop}(2006)}]{bishop2006pattern}%
  \BibitemOpen
  \bibfield  {author} {\bibinfo {author} {\bibfnamefont {C.~M.}\ \bibnamefont
  {Bishop}},\ }\href {https://cds.cern.ch/record/998831} {\emph {\bibinfo
  {title} {Pattern Recognition and Machine Learning}}}\ (\bibinfo  {publisher}
  {Springer, New York, NY},\ \bibinfo {year} {2006})\BibitemShut {NoStop}%
\bibitem [{\citenamefont {Goodfellow}\ \emph {et~al.}(2016)\citenamefont
  {Goodfellow}, \citenamefont {Bengio},\ and\ \citenamefont
  {Courville}}]{Goodfellow2016Deep}%
  \BibitemOpen
  \bibfield  {author} {\bibinfo {author} {\bibfnamefont {I.}~\bibnamefont
  {Goodfellow}}, \bibinfo {author} {\bibfnamefont {Y.}~\bibnamefont {Bengio}},\
  and\ \bibinfo {author} {\bibfnamefont {A.}~\bibnamefont {Courville}},\ }\href
  {https://www.deeplearningbook.org/front_matter.pdf} {\emph {\bibinfo {title}
  {Deep Learning}}}\ (\bibinfo  {publisher} {The MIT Press},\ \bibinfo {year}
  {2016})\BibitemShut {NoStop}%
\bibitem [{\citenamefont {Hinton}\ \emph {et~al.}(1984)\citenamefont {Hinton},
  \citenamefont {Sejnowski},\ and\ \citenamefont
  {Ackley}}]{hinton1984boltzmann}%
  \BibitemOpen
  \bibfield  {author} {\bibinfo {author} {\bibfnamefont {G.~E.}\ \bibnamefont
  {Hinton}}, \bibinfo {author} {\bibfnamefont {T.~J.}\ \bibnamefont
  {Sejnowski}},\ and\ \bibinfo {author} {\bibfnamefont {D.~H.}\ \bibnamefont
  {Ackley}},\ }\href {http://www.csri.utoronto.ca/~hinton/absps/bmtr.pdf}
  {\emph {\bibinfo {title} {Boltzmann machines: Constraint satisfaction
  networks that learn}}}\ (\bibinfo  {publisher} {Carnegie-Mellon University,
  Department of Computer Science Pittsburgh, PA},\ \bibinfo {year}
  {1984})\BibitemShut {NoStop}%
\bibitem [{\citenamefont {Smolensky}(1986)}]{smolensky1986information}%
  \BibitemOpen
  \bibfield  {author} {\bibinfo {author} {\bibfnamefont {P.}~\bibnamefont
  {Smolensky}},\ }\bibfield  {title} {\bibinfo {title} {Information processing
  in dynamical systems: Foundations of harmony theory},\ }in\ \href@noop {}
  {\emph {\bibinfo {booktitle} {Parallel Distributed Processing: Explorations
  in the Microstructure of Cognition, Vol. 1: Foundations}}}\ (\bibinfo
  {publisher} {MIT Press},\ \bibinfo {address} {Cambridge, MA, USA},\ \bibinfo
  {year} {1986})\ pp.\ \bibinfo {pages} {194--281}\BibitemShut {NoStop}%
\bibitem [{\citenamefont {Hinton}(2002)}]{hinton2002training}%
  \BibitemOpen
  \bibfield  {author} {\bibinfo {author} {\bibfnamefont {G.~E.}\ \bibnamefont
  {Hinton}},\ }\bibfield  {title} {\bibinfo {title} {Training products of
  experts by minimizing contrastive divergence},\ }\href
  {https://doi.org/10.1162/089976602760128018} {\bibfield  {journal} {\bibinfo
  {journal} {Neural Computation}\ }\textbf {\bibinfo {volume} {14}},\ \bibinfo
  {pages} {1771} (\bibinfo {year} {2002})}\BibitemShut {NoStop}%
\bibitem [{\citenamefont {Hinton}\ and\ \citenamefont
  {Salakhutdinov}(2006)}]{hinton2006reducing}%
  \BibitemOpen
  \bibfield  {author} {\bibinfo {author} {\bibfnamefont {G.~E.}\ \bibnamefont
  {Hinton}}\ and\ \bibinfo {author} {\bibfnamefont {R.~R.}\ \bibnamefont
  {Salakhutdinov}},\ }\bibfield  {title} {\bibinfo {title} {Reducing the
  dimensionality of data with neural networks},\ }\href
  {https://science.sciencemag.org/content/313/5786/504.abstract} {\bibfield
  {journal} {\bibinfo  {journal} {Science}\ }\textbf {\bibinfo {volume}
  {313}},\ \bibinfo {pages} {504} (\bibinfo {year} {2006})}\BibitemShut
  {NoStop}%
\bibitem [{\citenamefont {Han}\ \emph {et~al.}(2018)\citenamefont {Han},
  \citenamefont {Wang}, \citenamefont {Fan}, \citenamefont {Wang},\ and\
  \citenamefont {Zhang}}]{han2018unsupervised}%
  \BibitemOpen
  \bibfield  {author} {\bibinfo {author} {\bibfnamefont {Z.-Y.}\ \bibnamefont
  {Han}}, \bibinfo {author} {\bibfnamefont {J.}~\bibnamefont {Wang}}, \bibinfo
  {author} {\bibfnamefont {H.}~\bibnamefont {Fan}}, \bibinfo {author}
  {\bibfnamefont {L.}~\bibnamefont {Wang}},\ and\ \bibinfo {author}
  {\bibfnamefont {P.}~\bibnamefont {Zhang}},\ }\bibfield  {title} {\bibinfo
  {title} {Unsupervised generative modeling using matrix product states},\
  }\href {https://journals.aps.org/prx/abstract/10.1103/PhysRevX.8.031012}
  {\bibfield  {journal} {\bibinfo  {journal} {Physical Review X}\ }\textbf
  {\bibinfo {volume} {8}},\ \bibinfo {pages} {031012} (\bibinfo {year}
  {2018})}\BibitemShut {NoStop}%
\bibitem [{\citenamefont {Cheng}\ \emph {et~al.}(2018)\citenamefont {Cheng},
  \citenamefont {Chen},\ and\ \citenamefont {Wang}}]{cheng2018information}%
  \BibitemOpen
  \bibfield  {author} {\bibinfo {author} {\bibfnamefont {S.}~\bibnamefont
  {Cheng}}, \bibinfo {author} {\bibfnamefont {J.}~\bibnamefont {Chen}},\ and\
  \bibinfo {author} {\bibfnamefont {L.}~\bibnamefont {Wang}},\ }\bibfield
  {title} {\bibinfo {title} {Information perspective to probabilistic modeling:
  Boltzmann machines versus born machines},\ }\href
  {https://www.mdpi.com/1099-4300/20/8/583} {\bibfield  {journal} {\bibinfo
  {journal} {Entropy}\ }\textbf {\bibinfo {volume} {20}},\ \bibinfo {pages}
  {583} (\bibinfo {year} {2018})}\BibitemShut {NoStop}%
\bibitem [{\citenamefont {Cheng}\ \emph {et~al.}(2019)\citenamefont {Cheng},
  \citenamefont {Wang}, \citenamefont {Xiang},\ and\ \citenamefont
  {Zhang}}]{cheng2019tree}%
  \BibitemOpen
  \bibfield  {author} {\bibinfo {author} {\bibfnamefont {S.}~\bibnamefont
  {Cheng}}, \bibinfo {author} {\bibfnamefont {L.}~\bibnamefont {Wang}},
  \bibinfo {author} {\bibfnamefont {T.}~\bibnamefont {Xiang}},\ and\ \bibinfo
  {author} {\bibfnamefont {P.}~\bibnamefont {Zhang}},\ }\bibfield  {title}
  {\bibinfo {title} {Tree tensor networks for generative modeling},\ }\href
  {https://journals.aps.org/prb/abstract/10.1103/PhysRevB.99.155131} {\bibfield
   {journal} {\bibinfo  {journal} {Physical Review B}\ }\textbf {\bibinfo
  {volume} {99}},\ \bibinfo {pages} {155131} (\bibinfo {year}
  {2019})}\BibitemShut {NoStop}%
\bibitem [{\citenamefont {Glasser}\ \emph {et~al.}(2019)\citenamefont
  {Glasser}, \citenamefont {Sweke}, \citenamefont {Pancotti}, \citenamefont
  {Eisert},\ and\ \citenamefont {Cirac}}]{glasser2019expressive}%
  \BibitemOpen
  \bibfield  {author} {\bibinfo {author} {\bibfnamefont {I.}~\bibnamefont
  {Glasser}}, \bibinfo {author} {\bibfnamefont {R.}~\bibnamefont {Sweke}},
  \bibinfo {author} {\bibfnamefont {N.}~\bibnamefont {Pancotti}}, \bibinfo
  {author} {\bibfnamefont {J.}~\bibnamefont {Eisert}},\ and\ \bibinfo {author}
  {\bibfnamefont {I.}~\bibnamefont {Cirac}},\ }\bibfield  {title} {\bibinfo
  {title} {Expressive power of tensor-network factorizations for probabilistic
  modeling},\ }in\ \href
  {https://proceedings.neurips.cc/paper/2019/hash/b86e8d03fe992d1b0e19656875ee557c-Abstract.html}
  {\emph {\bibinfo {booktitle} {Advances in Neural Information Processing
  Systems}}}\ (\bibinfo {year} {2019})\ pp.\ \bibinfo {pages}
  {1498--1510}\BibitemShut {NoStop}%
\bibitem [{\citenamefont {Stoudenmire}\ and\ \citenamefont
  {Schwab}(2016)}]{Stoudenmire2016a}%
  \BibitemOpen
  \bibfield  {author} {\bibinfo {author} {\bibfnamefont {E.}~\bibnamefont
  {Stoudenmire}}\ and\ \bibinfo {author} {\bibfnamefont {D.~J.}\ \bibnamefont
  {Schwab}},\ }\bibfield  {title} {\bibinfo {title} {Supervised learning with
  tensor networks},\ }in\ \href
  {http://www.perimeterinstitute.ca/videos/learning-quantum-inspired-tensor-networks}
  {\emph {\bibinfo {booktitle} {Advances In Neural Information Processing
  Systems}}}\ (\bibinfo {year} {2016})\ pp.\ \bibinfo {pages}
  {4799--4807}\BibitemShut {NoStop}%
\bibitem [{\citenamefont {Stoudenmire}(2018)}]{stoudenmire2018learning}%
  \BibitemOpen
  \bibfield  {author} {\bibinfo {author} {\bibfnamefont {E.~M.}\ \bibnamefont
  {Stoudenmire}},\ }\bibfield  {title} {\bibinfo {title} {Learning relevant
  features of data with multi-scale tensor networks},\ }\href
  {https://doi.org/10.1088/2058-9565/aaba1a} {\bibfield  {journal} {\bibinfo
  {journal} {Quantum Science and Technology}\ }\textbf {\bibinfo {volume}
  {3}},\ \bibinfo {pages} {034003} (\bibinfo {year} {2018})}\BibitemShut
  {NoStop}%
\bibitem [{\citenamefont {Cheng}\ \emph {et~al.}(2021)\citenamefont {Cheng},
  \citenamefont {Wang},\ and\ \citenamefont {Zhang}}]{cheng2021supervised}%
  \BibitemOpen
  \bibfield  {author} {\bibinfo {author} {\bibfnamefont {S.}~\bibnamefont
  {Cheng}}, \bibinfo {author} {\bibfnamefont {L.}~\bibnamefont {Wang}},\ and\
  \bibinfo {author} {\bibfnamefont {P.}~\bibnamefont {Zhang}},\ }\bibfield
  {title} {\bibinfo {title} {Supervised learning with projected entangled pair
  states},\ }\href
  {https://journals.aps.org/prb/abstract/10.1103/PhysRevB.103.125117}
  {\bibfield  {journal} {\bibinfo  {journal} {Physical Review B}\ }\textbf
  {\bibinfo {volume} {103}},\ \bibinfo {pages} {125117} (\bibinfo {year}
  {2021})}\BibitemShut {NoStop}%
\bibitem [{\citenamefont {Meng}\ \emph {et~al.}(2020)\citenamefont {Meng},
  \citenamefont {Zhang}, \citenamefont {Zhang}, \citenamefont {Gao},\ and\
  \citenamefont {Ran}}]{meng2020residual}%
  \BibitemOpen
  \bibfield  {author} {\bibinfo {author} {\bibfnamefont {Y.-M.}\ \bibnamefont
  {Meng}}, \bibinfo {author} {\bibfnamefont {J.}~\bibnamefont {Zhang}},
  \bibinfo {author} {\bibfnamefont {P.}~\bibnamefont {Zhang}}, \bibinfo
  {author} {\bibfnamefont {C.}~\bibnamefont {Gao}},\ and\ \bibinfo {author}
  {\bibfnamefont {S.-J.}\ \bibnamefont {Ran}},\ }\bibfield  {title} {\bibinfo
  {title} {Residual matrix product state for machine learning},\ }\href
  {http://arxiv.org/abs/2012.11841v1} {\bibfield  {journal} {\bibinfo
  {journal} {arxiv:2012.11841}\ } (\bibinfo {year} {2020})}\BibitemShut
  {NoStop}%
\bibitem [{\citenamefont {Liu}\ \emph {et~al.}(2019)\citenamefont {Liu},
  \citenamefont {Ran}, \citenamefont {Wittek}, \citenamefont {Peng},
  \citenamefont {Garc{\'\i}a}, \citenamefont {Su},\ and\ \citenamefont
  {Lewenstein}}]{liu2019machine}%
  \BibitemOpen
  \bibfield  {author} {\bibinfo {author} {\bibfnamefont {D.}~\bibnamefont
  {Liu}}, \bibinfo {author} {\bibfnamefont {S.-J.}\ \bibnamefont {Ran}},
  \bibinfo {author} {\bibfnamefont {P.}~\bibnamefont {Wittek}}, \bibinfo
  {author} {\bibfnamefont {C.}~\bibnamefont {Peng}}, \bibinfo {author}
  {\bibfnamefont {R.~B.}\ \bibnamefont {Garc{\'\i}a}}, \bibinfo {author}
  {\bibfnamefont {G.}~\bibnamefont {Su}},\ and\ \bibinfo {author}
  {\bibfnamefont {M.}~\bibnamefont {Lewenstein}},\ }\bibfield  {title}
  {\bibinfo {title} {Machine learning by unitary tensor network of hierarchical
  tree structure},\ }\href
  {https://iopscience.iop.org/article/10.1088/1367-2630/ab31ef/meta} {\bibfield
   {journal} {\bibinfo  {journal} {New Journal of Physics}\ }\textbf {\bibinfo
  {volume} {21}},\ \bibinfo {pages} {073059} (\bibinfo {year}
  {2019})}\BibitemShut {NoStop}%
\bibitem [{\citenamefont {{Perez-Garcia}}\ \emph {et~al.}(2007)\citenamefont
  {{Perez-Garcia}}, \citenamefont {Verstraete}, \citenamefont {Wolf},\ and\
  \citenamefont {Cirac}}]{perez2006matrix}%
  \BibitemOpen
  \bibfield  {author} {\bibinfo {author} {\bibfnamefont {D.}~\bibnamefont
  {{Perez-Garcia}}}, \bibinfo {author} {\bibfnamefont {F.}~\bibnamefont
  {Verstraete}}, \bibinfo {author} {\bibfnamefont {M.~M.}\ \bibnamefont
  {Wolf}},\ and\ \bibinfo {author} {\bibfnamefont {J.~I.}\ \bibnamefont
  {Cirac}},\ }\bibfield  {title} {\bibinfo {title} {Matrix product state
  representations},\ }\href@noop {} {\bibfield  {journal} {\bibinfo  {journal}
  {Quantum Info. Comput.}\ }\textbf {\bibinfo {volume} {7}},\ \bibinfo {pages}
  {401} (\bibinfo {year} {2007})}\BibitemShut {NoStop}%
\bibitem [{\citenamefont {Shi}\ \emph {et~al.}(2006)\citenamefont {Shi},
  \citenamefont {Duan},\ and\ \citenamefont {Vidal}}]{2006Classical}%
  \BibitemOpen
  \bibfield  {author} {\bibinfo {author} {\bibfnamefont {Y.}~\bibnamefont
  {Shi}}, \bibinfo {author} {\bibfnamefont {L.}~\bibnamefont {Duan}},\ and\
  \bibinfo {author} {\bibfnamefont {G.}~\bibnamefont {Vidal}},\ }\bibfield
  {title} {\bibinfo {title} {Classical simulation of quantum many-body systems
  with a tree tensor network},\ }\href
  {https://journals.aps.org/pra/abstract/10.1103/PhysRevA.74.022320} {\bibfield
   {journal} {\bibinfo  {journal} {Physical Review A}\ }\textbf {\bibinfo
  {volume} {74}},\ \bibinfo {pages} {154} (\bibinfo {year} {2006})}\BibitemShut
  {NoStop}%
\bibitem [{\citenamefont {Li}\ \emph {et~al.}(2021)\citenamefont {Li},
  \citenamefont {Pan}, \citenamefont {Zhou},\ and\ \citenamefont
  {Zhang}}]{li2021boltzmann}%
  \BibitemOpen
  \bibfield  {author} {\bibinfo {author} {\bibfnamefont {S.}~\bibnamefont
  {Li}}, \bibinfo {author} {\bibfnamefont {F.}~\bibnamefont {Pan}}, \bibinfo
  {author} {\bibfnamefont {P.}~\bibnamefont {Zhou}},\ and\ \bibinfo {author}
  {\bibfnamefont {P.}~\bibnamefont {Zhang}},\ }\bibfield  {title} {\bibinfo
  {title} {Boltzmann machines as two-dimensional tensor networks},\ }\href
  {https://doi.org/10.1103/PhysRevB.104.075154} {\bibfield  {journal} {\bibinfo
   {journal} {Phys. Rev. B}\ }\textbf {\bibinfo {volume} {104}},\ \bibinfo
  {pages} {075154} (\bibinfo {year} {2021})}\BibitemShut {NoStop}%
\bibitem [{\citenamefont {Wu}\ \emph {et~al.}(2019)\citenamefont {Wu},
  \citenamefont {Wang},\ and\ \citenamefont {Zhang}}]{wu2019solving}%
  \BibitemOpen
  \bibfield  {author} {\bibinfo {author} {\bibfnamefont {D.}~\bibnamefont
  {Wu}}, \bibinfo {author} {\bibfnamefont {L.}~\bibnamefont {Wang}},\ and\
  \bibinfo {author} {\bibfnamefont {P.}~\bibnamefont {Zhang}},\ }\bibfield
  {title} {\bibinfo {title} {Solving statistical mechanics using variational
  autoregressive networks},\ }\href
  {https://journals.aps.org/prl/abstract/10.1103/PhysRevLett.122.080602}
  {\bibfield  {journal} {\bibinfo  {journal} {Physical Review Letters}\
  }\textbf {\bibinfo {volume} {122}},\ \bibinfo {pages} {080602} (\bibinfo
  {year} {2019})}\BibitemShut {NoStop}%
\bibitem [{\citenamefont {Williams}(1992)}]{williams1992simple}%
  \BibitemOpen
  \bibfield  {author} {\bibinfo {author} {\bibfnamefont {R.~J.}\ \bibnamefont
  {Williams}},\ }\bibfield  {title} {\bibinfo {title} {Simple statistical
  gradient-following algorithms for connectionist reinforcement learning},\
  }\href {https://doi.org/10.1007/BF00992696} {\bibfield  {journal} {\bibinfo
  {journal} {Machine Learning}\ }\textbf {\bibinfo {volume} {8}},\ \bibinfo
  {pages} {229} (\bibinfo {year} {1992})}\BibitemShut {NoStop}%
\bibitem [{\citenamefont {Cichocki}\ \emph {et~al.}(2016)\citenamefont
  {Cichocki}, \citenamefont {Lee}, \citenamefont {Oseledets}, \citenamefont
  {Phan}, \citenamefont {Zhao},\ and\ \citenamefont {Mandic}}]{MAL-059}%
  \BibitemOpen
  \bibfield  {author} {\bibinfo {author} {\bibfnamefont {A.}~\bibnamefont
  {Cichocki}}, \bibinfo {author} {\bibfnamefont {N.}~\bibnamefont {Lee}},
  \bibinfo {author} {\bibfnamefont {I.}~\bibnamefont {Oseledets}}, \bibinfo
  {author} {\bibfnamefont {A.-H.}\ \bibnamefont {Phan}}, \bibinfo {author}
  {\bibfnamefont {Q.}~\bibnamefont {Zhao}},\ and\ \bibinfo {author}
  {\bibfnamefont {D.~P.}\ \bibnamefont {Mandic}},\ }\bibfield  {title}
  {\bibinfo {title} {Tensor networks for dimensionality reduction and
  large-scale optimization: Part 1 low-rank tensor decompositions},\ }\href
  {https://doi.org/10.1561/2200000059} {\bibfield  {journal} {\bibinfo
  {journal} {Foundations and Trends\textregistered\ in Machine Learning}\
  }\textbf {\bibinfo {volume} {9}},\ \bibinfo {pages} {249} (\bibinfo {year}
  {2016})}\BibitemShut {NoStop}%
\bibitem [{\citenamefont {Clark}(2018)}]{clark2018unifying}%
  \BibitemOpen
  \bibfield  {author} {\bibinfo {author} {\bibfnamefont {S.~R.}\ \bibnamefont
  {Clark}},\ }\bibfield  {title} {\bibinfo {title} {Unifying neural-network
  quantum states and correlator product states via tensor networks},\ }\href
  {https://iopscience.iop.org/article/10.1088/1751-8121/aaaaf2/meta?casa_token=Ytmy6mx6VGMAAAAA:UFaK6UEJoJZIWRk4fe4pJlFHMY6QCjQteIU3O3jszWs4cU5b_YWhEgUEPQVs4v9bqQNt1tW28vMLfw}
  {\bibfield  {journal} {\bibinfo  {journal} {Journal of Physics A:
  Mathematical and Theoretical}\ }\textbf {\bibinfo {volume} {51}},\ \bibinfo
  {pages} {135301} (\bibinfo {year} {2018})}\BibitemShut {NoStop}%
\bibitem [{sup()}]{supplemental_material}%
  \BibitemOpen
  \href@noop {} {}\bibinfo {note} {See Supplemental Material for a detailed
  description of the autoregressive matrix product states model and additional
  results on the task of generative modeling and reinforcement learning in
  statistical physics, which includes Refs.~[24-30].}\BibitemShut {Stop}%
\bibitem [{\citenamefont {Oseledets}(2011)}]{oseledets2011tensor}%
  \BibitemOpen
  \bibfield  {author} {\bibinfo {author} {\bibfnamefont {I.~V.}\ \bibnamefont
  {Oseledets}},\ }\bibfield  {title} {\bibinfo {title} {Tensor-train
  decomposition},\ }\href {https://epubs.siam.org/doi/abs/10.1137/090752286}
  {\bibfield  {journal} {\bibinfo  {journal} {SIAM Journal on Scientific
  Computing}\ }\textbf {\bibinfo {volume} {33}},\ \bibinfo {pages} {2295}
  (\bibinfo {year} {2011})}\BibitemShut {NoStop}%
\bibitem [{\citenamefont {Glasser}\ \emph {et~al.}(2018)\citenamefont
  {Glasser}, \citenamefont {Pancotti}, \citenamefont {August}, \citenamefont
  {Rodriguez},\ and\ \citenamefont {Cirac}}]{glasser2018neural}%
  \BibitemOpen
  \bibfield  {author} {\bibinfo {author} {\bibfnamefont {I.}~\bibnamefont
  {Glasser}}, \bibinfo {author} {\bibfnamefont {N.}~\bibnamefont {Pancotti}},
  \bibinfo {author} {\bibfnamefont {M.}~\bibnamefont {August}}, \bibinfo
  {author} {\bibfnamefont {I.~D.}\ \bibnamefont {Rodriguez}},\ and\ \bibinfo
  {author} {\bibfnamefont {J.~I.}\ \bibnamefont {Cirac}},\ }\bibfield  {title}
  {\bibinfo {title} {Neural-network quantum states, string-bond states, and
  chiral topological states},\ }\href
  {https://journals.aps.org/prx/abstract/10.1103/PhysRevX.8.011006} {\bibfield
  {journal} {\bibinfo  {journal} {Physical Review X}\ }\textbf {\bibinfo
  {volume} {8}},\ \bibinfo {pages} {011006} (\bibinfo {year}
  {2018})}\BibitemShut {NoStop}%
\bibitem [{\citenamefont {Kingma}\ and\ \citenamefont
  {Ba}(2015)}]{kingma2014adam}%
  \BibitemOpen
  \bibfield  {author} {\bibinfo {author} {\bibfnamefont {D.~P.}\ \bibnamefont
  {Kingma}}\ and\ \bibinfo {author} {\bibfnamefont {J.}~\bibnamefont {Ba}},\
  }\bibfield  {title} {\bibinfo {title} {Adam: {A} method for stochastic
  optimization},\ }in\ \href {http://arxiv.org/abs/1412.6980} {\emph {\bibinfo
  {booktitle} {3rd International Conference on Learning Representations, {ICLR}
  2015, San Diego, CA, USA, May 7-9, 2015, Conference Track Proceedings}}},\
  \bibinfo {editor} {edited by\ \bibinfo {editor} {\bibfnamefont
  {Y.}~\bibnamefont {Bengio}}\ and\ \bibinfo {editor} {\bibfnamefont
  {Y.}~\bibnamefont {LeCun}}}\ (\bibinfo {year} {2015})\BibitemShut {NoStop}%
\bibitem [{\citenamefont {M{\"u}ller}\ \emph {et~al.}(2007)\citenamefont
  {M{\"u}ller}, \citenamefont {Studer},\ and\ \citenamefont
  {Ritschard}}]{muller2007classification}%
  \BibitemOpen
  \bibfield  {author} {\bibinfo {author} {\bibfnamefont {N.~S.}\ \bibnamefont
  {M{\"u}ller}}, \bibinfo {author} {\bibfnamefont {M.}~\bibnamefont {Studer}},\
  and\ \bibinfo {author} {\bibfnamefont {G.}~\bibnamefont {Ritschard}},\
  }\bibfield  {title} {\bibinfo {title} {Classification de parcours de vie
  {\`a} l'aide de l'optimal matching},\ }\href
  {https://archive-ouverte.unige.ch/unige:4532} {\bibfield  {journal} {\bibinfo
   {journal} {XIVe Rencontre de la Soci{\'e}t{\'e} francophone de
  classification (SFC 2007)}\ ,\ \bibinfo {pages} {157}} (\bibinfo {year}
  {2007})}\BibitemShut {NoStop}%
\bibitem [{\citenamefont {Dua}\ and\ \citenamefont {Graff}(2017)}]{Dua:2019}%
  \BibitemOpen
  \bibfield  {author} {\bibinfo {author} {\bibfnamefont {D.}~\bibnamefont
  {Dua}}\ and\ \bibinfo {author} {\bibfnamefont {C.}~\bibnamefont {Graff}},\
  }\href {http://archive.ics.uci.edu/ml} {\bibinfo {title} {{UCI} machine
  learning repository}} (\bibinfo {year} {2017})\BibitemShut {NoStop}%
\bibitem [{dow()}]{download}%
  \BibitemOpen
  \href@noop {} {}\bibinfo {note} {\url
  {https://github.com/glivan/tensor\_networks\_for\_probabilistic\_modeling}
  (Accessed on 1-October-2020)}\BibitemShut {NoStop}%
\bibitem [{\citenamefont {Nicoli}\ \emph {et~al.}(2020)\citenamefont {Nicoli},
  \citenamefont {Nakajima}, \citenamefont {Strodthoff}, \citenamefont {Samek},
  \citenamefont {M{\"u}ller},\ and\ \citenamefont
  {Kessel}}]{nicoli2020asymptotically}%
  \BibitemOpen
  \bibfield  {author} {\bibinfo {author} {\bibfnamefont {K.~A.}\ \bibnamefont
  {Nicoli}}, \bibinfo {author} {\bibfnamefont {S.}~\bibnamefont {Nakajima}},
  \bibinfo {author} {\bibfnamefont {N.}~\bibnamefont {Strodthoff}}, \bibinfo
  {author} {\bibfnamefont {W.}~\bibnamefont {Samek}}, \bibinfo {author}
  {\bibfnamefont {K.-R.}\ \bibnamefont {M{\"u}ller}},\ and\ \bibinfo {author}
  {\bibfnamefont {P.}~\bibnamefont {Kessel}},\ }\bibfield  {title} {\bibinfo
  {title} {Asymptotically unbiased estimation of physical observables with
  neural samplers},\ }\href
  {https://journals.aps.org/pre/abstract/10.1103/PhysRevE.101.023304}
  {\bibfield  {journal} {\bibinfo  {journal} {Physical Review E}\ }\textbf
  {\bibinfo {volume} {101}},\ \bibinfo {pages} {023304} (\bibinfo {year}
  {2020})}\BibitemShut {NoStop}%
\bibitem [{lym()}]{lymphodataset}%
  \BibitemOpen
  \href@noop {} {}\bibinfo {note} {\url
  {http://archive.ics.uci.edu/ml/datasets/Lymphography?ref=datanews.io}
  (Accessed on 1-October-2020)}\BibitemShut {NoStop}%
\bibitem [{\citenamefont {LeCun}\ \emph {et~al.}(2015)\citenamefont {LeCun},
  \citenamefont {Bengio},\ and\ \citenamefont {Hinton}}]{lecun2015deep}%
  \BibitemOpen
  \bibfield  {author} {\bibinfo {author} {\bibfnamefont {Y.}~\bibnamefont
  {LeCun}}, \bibinfo {author} {\bibfnamefont {Y.}~\bibnamefont {Bengio}},\ and\
  \bibinfo {author} {\bibfnamefont {G.}~\bibnamefont {Hinton}},\ }\bibfield
  {title} {\bibinfo {title} {Deep learning},\ }\href
  {https://www.nature.com/articles/nature14539} {\bibfield  {journal} {\bibinfo
   {journal} {Nature}\ }\textbf {\bibinfo {volume} {521}},\ \bibinfo {pages}
  {436} (\bibinfo {year} {2015})}\BibitemShut {NoStop}%
\bibitem [{\citenamefont {Hopfield}(1982)}]{hopfield1982neural}%
  \BibitemOpen
  \bibfield  {author} {\bibinfo {author} {\bibfnamefont {J.~J.}\ \bibnamefont
  {Hopfield}},\ }\bibfield  {title} {\bibinfo {title} {Neural networks and
  physical systems with emergent collective computational abilities},\ }\href
  {https://www.pnas.org/content/79/8/2554.short} {\bibfield  {journal}
  {\bibinfo  {journal} {Proceedings of the national academy of sciences}\
  }\textbf {\bibinfo {volume} {79}},\ \bibinfo {pages} {2554} (\bibinfo {year}
  {1982})}\BibitemShut {NoStop}%
\bibitem [{\citenamefont {Salakhutdinov}\ and\ \citenamefont
  {Murray}(2008)}]{salakhutdinov2008quantitative}%
  \BibitemOpen
  \bibfield  {author} {\bibinfo {author} {\bibfnamefont {R.}~\bibnamefont
  {Salakhutdinov}}\ and\ \bibinfo {author} {\bibfnamefont {I.}~\bibnamefont
  {Murray}},\ }\bibfield  {title} {\bibinfo {title} {On the quantitative
  analysis of deep belief networks},\ }in\ \href
  {https://doi.org/10.1145/1390156.1390266} {\emph {\bibinfo {booktitle}
  {Machine Learning, Proceedings of the Twenty-Fifth International Conference
  ({{ICML}} 2008), Helsinki, Finland, June 5-9, 2008}}},\ \bibinfo {series}
  {{{ACM}} International Conference Proceeding Series}, Vol.\ \bibinfo {volume}
  {307},\ \bibinfo {editor} {edited by\ \bibinfo {editor} {\bibfnamefont
  {W.~W.}\ \bibnamefont {Cohen}}, \bibinfo {editor} {\bibfnamefont
  {A.}~\bibnamefont {McCallum}},\ and\ \bibinfo {editor} {\bibfnamefont
  {S.~T.}\ \bibnamefont {Roweis}}}\ (\bibinfo  {publisher} {{ACM}},\ \bibinfo
  {year} {2008})\ pp.\ \bibinfo {pages} {872--879}\BibitemShut {NoStop}%
\bibitem [{\citenamefont {Uria}\ \emph {et~al.}(2014)\citenamefont {Uria},
  \citenamefont {Murray},\ and\ \citenamefont {Larochelle}}]{uria2014deep}%
  \BibitemOpen
  \bibfield  {author} {\bibinfo {author} {\bibfnamefont {B.}~\bibnamefont
  {Uria}}, \bibinfo {author} {\bibfnamefont {I.}~\bibnamefont {Murray}},\ and\
  \bibinfo {author} {\bibfnamefont {H.}~\bibnamefont {Larochelle}},\ }\bibfield
   {title} {\bibinfo {title} {A deep and tractable density estimator},\ }in\
  \href@noop {} {\emph {\bibinfo {booktitle} {Proceedings of the 31th
  International Conference on Machine Learning, {{ICML}} 2014, Beijing, China,
  21-26 June 2014}}},\ \bibinfo {series} {{{JMLR}} Workshop and Conference
  Proceedings}, Vol.~\bibinfo {volume} {32}\ (\bibinfo  {publisher}
  {{JMLR.org}},\ \bibinfo {year} {2014})\ pp.\ \bibinfo {pages}
  {467--475}\BibitemShut {NoStop}%
\bibitem [{\citenamefont {Germain}\ \emph {et~al.}(2015)\citenamefont
  {Germain}, \citenamefont {Gregor}, \citenamefont {Murray},\ and\
  \citenamefont {Larochelle}}]{germain2015made}%
  \BibitemOpen
  \bibfield  {author} {\bibinfo {author} {\bibfnamefont {M.}~\bibnamefont
  {Germain}}, \bibinfo {author} {\bibfnamefont {K.}~\bibnamefont {Gregor}},
  \bibinfo {author} {\bibfnamefont {I.}~\bibnamefont {Murray}},\ and\ \bibinfo
  {author} {\bibfnamefont {H.}~\bibnamefont {Larochelle}},\ }\bibfield  {title}
  {\bibinfo {title} {{{MADE}}: {{Masked}} autoencoder for distribution
  estimation},\ }in\ \href@noop {} {\emph {\bibinfo {booktitle} {Proceedings of
  the 32nd International Conference on Machine Learning, {{ICML}} 2015, Lille,
  France, 6-11 July 2015}}},\ \bibinfo {series} {{{JMLR}} Workshop and
  Conference Proceedings}, Vol.~\bibinfo {volume} {37},\ \bibinfo {editor}
  {edited by\ \bibinfo {editor} {\bibfnamefont {F.~R.}\ \bibnamefont {Bach}}\
  and\ \bibinfo {editor} {\bibfnamefont {D.~M.}\ \bibnamefont {Blei}}}\
  (\bibinfo  {publisher} {{JMLR.org}},\ \bibinfo {year} {2015})\ pp.\ \bibinfo
  {pages} {881--889}\BibitemShut {NoStop}%
\bibitem [{\citenamefont {{van den Oord}}\ \emph {et~al.}(2016)\citenamefont
  {{van den Oord}}, \citenamefont {Kalchbrenner},\ and\ \citenamefont
  {Kavukcuoglu}}]{van2016pixel}%
  \BibitemOpen
  \bibfield  {author} {\bibinfo {author} {\bibfnamefont {A.}~\bibnamefont {{van
  den Oord}}}, \bibinfo {author} {\bibfnamefont {N.}~\bibnamefont
  {Kalchbrenner}},\ and\ \bibinfo {author} {\bibfnamefont {K.}~\bibnamefont
  {Kavukcuoglu}},\ }\bibfield  {title} {\bibinfo {title} {Pixel recurrent
  neural networks},\ }in\ \href@noop {} {\emph {\bibinfo {booktitle}
  {Proceedings of the 33nd International Conference on Machine Learning,
  {{ICML}} 2016, New York City, {{NY}}, {{USA}}, June 19-24, 2016}}},\ \bibinfo
  {series} {{{JMLR}} Workshop and Conference Proceedings}, Vol.~\bibinfo
  {volume} {48},\ \bibinfo {editor} {edited by\ \bibinfo {editor}
  {\bibfnamefont {M.-F.}\ \bibnamefont {Balcan}}\ and\ \bibinfo {editor}
  {\bibfnamefont {K.~Q.}\ \bibnamefont {Weinberger}}}\ (\bibinfo  {publisher}
  {{JMLR.org}},\ \bibinfo {year} {2016})\ pp.\ \bibinfo {pages}
  {1747--1756}\BibitemShut {NoStop}%
\bibitem [{bin()}]{binarizedMNIST}%
  \BibitemOpen
  \href@noop {} {\bibinfo {title} {\url
  {http://www.dmi.usherb.ca/~larocheh/mlpython/\_modules/datasets/binarized\_mnist.html}
  (accessed on 1-october-2020)}}\BibitemShut {NoStop}%
\bibitem [{\citenamefont {Larochelle}\ and\ \citenamefont
  {Murray}(2011)}]{larochelle2011neural}%
  \BibitemOpen
  \bibfield  {author} {\bibinfo {author} {\bibfnamefont {H.}~\bibnamefont
  {Larochelle}}\ and\ \bibinfo {author} {\bibfnamefont {I.}~\bibnamefont
  {Murray}},\ }\bibfield  {title} {\bibinfo {title} {The neural autoregressive
  distribution estimator},\ }in\ \href@noop {} {\emph {\bibinfo {booktitle}
  {Proceedings of the Fourteenth International Conference on Artificial
  Intelligence and Statistics, {{AISTATS}} 2011, Fort Lauderdale, {{USA}},
  April 11-13, 2011}}},\ \bibinfo {series} {{{JMLR}} Proceedings},
  Vol.~\bibinfo {volume} {15},\ \bibinfo {editor} {edited by\ \bibinfo {editor}
  {\bibfnamefont {G.~J.}\ \bibnamefont {Gordon}}, \bibinfo {editor}
  {\bibfnamefont {D.~B.}\ \bibnamefont {Dunson}},\ and\ \bibinfo {editor}
  {\bibfnamefont {M.}~\bibnamefont {Dud{\'i}k}}}\ (\bibinfo  {publisher}
  {{JMLR.org}},\ \bibinfo {year} {2011})\ pp.\ \bibinfo {pages}
  {29--37}\BibitemShut {NoStop}%
\bibitem [{\citenamefont {Sutton}\ and\ \citenamefont
  {Barto}(2018)}]{sutton2018reinforcement}%
  \BibitemOpen
  \bibfield  {author} {\bibinfo {author} {\bibfnamefont {R.~S.}\ \bibnamefont
  {Sutton}}\ and\ \bibinfo {author} {\bibfnamefont {A.~G.}\ \bibnamefont
  {Barto}},\ }\href
  {https://books.google.com/books?hl=zh-CN&lr=&id=uWV0DwAAQBAJ&oi=fnd&pg=PR7&ots=minLm6-Xl0&sig=7zNumAON-vSegLaoH7Jzbmk7Jvc#v=onepage&q&f=false}
  {\emph {\bibinfo {title} {Reinforcement learning: An introduction}}}\
  (\bibinfo  {publisher} {MIT press},\ \bibinfo {year} {2018})\BibitemShut
  {NoStop}%
\bibitem [{\citenamefont {Silver}\ \emph {et~al.}(2017)\citenamefont {Silver},
  \citenamefont {Schrittwieser}, \citenamefont {Simonyan}, \citenamefont
  {Antonoglou}, \citenamefont {Huang}, \citenamefont {Guez}, \citenamefont
  {Hubert}, \citenamefont {Baker}, \citenamefont {Lai}, \citenamefont {Bolton}
  \emph {et~al.}}]{silver2017mastering}%
  \BibitemOpen
  \bibfield  {author} {\bibinfo {author} {\bibfnamefont {D.}~\bibnamefont
  {Silver}}, \bibinfo {author} {\bibfnamefont {J.}~\bibnamefont
  {Schrittwieser}}, \bibinfo {author} {\bibfnamefont {K.}~\bibnamefont
  {Simonyan}}, \bibinfo {author} {\bibfnamefont {I.}~\bibnamefont
  {Antonoglou}}, \bibinfo {author} {\bibfnamefont {A.}~\bibnamefont {Huang}},
  \bibinfo {author} {\bibfnamefont {A.}~\bibnamefont {Guez}}, \bibinfo {author}
  {\bibfnamefont {T.}~\bibnamefont {Hubert}}, \bibinfo {author} {\bibfnamefont
  {L.}~\bibnamefont {Baker}}, \bibinfo {author} {\bibfnamefont
  {M.}~\bibnamefont {Lai}}, \bibinfo {author} {\bibfnamefont {A.}~\bibnamefont
  {Bolton}}, \emph {et~al.},\ }\bibfield  {title} {\bibinfo {title} {Mastering
  the game of go without human knowledge},\ }\href
  {https://www.nature.com/articles/nature24270?sf123103138=1} {\bibfield
  {journal} {\bibinfo  {journal} {nature}\ }\textbf {\bibinfo {volume} {550}},\
  \bibinfo {pages} {354} (\bibinfo {year} {2017})}\BibitemShut {NoStop}%
\bibitem [{\citenamefont {Pan}\ \emph {et~al.}(2021)\citenamefont {Pan},
  \citenamefont {Zhou}, \citenamefont {Zhou},\ and\ \citenamefont
  {Zhang}}]{pan2019solving}%
  \BibitemOpen
  \bibfield  {author} {\bibinfo {author} {\bibfnamefont {F.}~\bibnamefont
  {Pan}}, \bibinfo {author} {\bibfnamefont {P.}~\bibnamefont {Zhou}}, \bibinfo
  {author} {\bibfnamefont {H.-J.}\ \bibnamefont {Zhou}},\ and\ \bibinfo
  {author} {\bibfnamefont {P.}~\bibnamefont {Zhang}},\ }\bibfield  {title}
  {\bibinfo {title} {Solving statistical mechanics on sparse graphs with
  feedback-set variational autoregressive networks},\ }\href
  {https://doi.org/10.1103/PhysRevE.103.012103} {\bibfield  {journal} {\bibinfo
   {journal} {Phys. Rev. E}\ }\textbf {\bibinfo {volume} {103}},\ \bibinfo
  {pages} {012103} (\bibinfo {year} {2021})}\BibitemShut {NoStop}%
\bibitem [{\citenamefont {Carleo}\ and\ \citenamefont
  {Troyer}(2017)}]{carleo2017solving}%
  \BibitemOpen
  \bibfield  {author} {\bibinfo {author} {\bibfnamefont {G.}~\bibnamefont
  {Carleo}}\ and\ \bibinfo {author} {\bibfnamefont {M.}~\bibnamefont
  {Troyer}},\ }\bibfield  {title} {\bibinfo {title} {Solving the quantum
  many-body problem with artificial neural networks},\ }\href
  {https://science.sciencemag.org/content/355/6325/602.abstract} {\bibfield
  {journal} {\bibinfo  {journal} {Science}\ }\textbf {\bibinfo {volume}
  {355}},\ \bibinfo {pages} {602} (\bibinfo {year} {2017})}\BibitemShut
  {NoStop}%
\bibitem [{\citenamefont {Sharir}\ \emph {et~al.}(2020)\citenamefont {Sharir},
  \citenamefont {Levine}, \citenamefont {Wies}, \citenamefont {Carleo},\ and\
  \citenamefont {Shashua}}]{sharir2020deep}%
  \BibitemOpen
  \bibfield  {author} {\bibinfo {author} {\bibfnamefont {O.}~\bibnamefont
  {Sharir}}, \bibinfo {author} {\bibfnamefont {Y.}~\bibnamefont {Levine}},
  \bibinfo {author} {\bibfnamefont {N.}~\bibnamefont {Wies}}, \bibinfo {author}
  {\bibfnamefont {G.}~\bibnamefont {Carleo}},\ and\ \bibinfo {author}
  {\bibfnamefont {A.}~\bibnamefont {Shashua}},\ }\bibfield  {title} {\bibinfo
  {title} {Deep autoregressive models for the efficient variational simulation
  of many-body quantum systems},\ }\href
  {https://journals.aps.org/prl/abstract/10.1103/PhysRevLett.124.020503}
  {\bibfield  {journal} {\bibinfo  {journal} {Physical Review Letters}\
  }\textbf {\bibinfo {volume} {124}},\ \bibinfo {pages} {020503} (\bibinfo
  {year} {2020})}\BibitemShut {NoStop}%
\bibitem [{\citenamefont {Sherrington}\ and\ \citenamefont
  {Kirkpatrick}(1975)}]{sherrington1975solvable}%
  \BibitemOpen
  \bibfield  {author} {\bibinfo {author} {\bibfnamefont {D.}~\bibnamefont
  {Sherrington}}\ and\ \bibinfo {author} {\bibfnamefont {S.}~\bibnamefont
  {Kirkpatrick}},\ }\bibfield  {title} {\bibinfo {title} {Solvable model of a
  spin-glass},\ }\href
  {https://journals.aps.org/prl/abstract/10.1103/PhysRevLett.35.1792}
  {\bibfield  {journal} {\bibinfo  {journal} {Physical Review Letters}\
  }\textbf {\bibinfo {volume} {35}},\ \bibinfo {pages} {1792} (\bibinfo {year}
  {1975})}\BibitemShut {NoStop}%
\bibitem [{\citenamefont {Anderson}\ and\ \citenamefont
  {Peterson}(1987)}]{anderson1987mean}%
  \BibitemOpen
  \bibfield  {author} {\bibinfo {author} {\bibfnamefont {J.~R.}\ \bibnamefont
  {Anderson}}\ and\ \bibinfo {author} {\bibfnamefont {C.}~\bibnamefont
  {Peterson}},\ }\bibfield  {title} {\bibinfo {title} {A mean field theory
  learning algorithm for neural networks},\ }\href
  {http://home.thep.lu.se/~carsten/pubs/lu_tp_87_01.pdf} {\bibfield  {journal}
  {\bibinfo  {journal} {Complex Systems}\ }\textbf {\bibinfo {volume} {1}},\
  \bibinfo {pages} {995} (\bibinfo {year} {1987})}\BibitemShut {NoStop}%
\bibitem [{\citenamefont {Boixo}\ \emph {et~al.}(2018)\citenamefont {Boixo},
  \citenamefont {Isakov}, \citenamefont {Smelyanskiy}, \citenamefont {Babbush},
  \citenamefont {Ding}, \citenamefont {Jiang}, \citenamefont {Bremner},
  \citenamefont {Martinis},\ and\ \citenamefont
  {Neven}}]{boixo2018characterizing}%
  \BibitemOpen
  \bibfield  {author} {\bibinfo {author} {\bibfnamefont {S.}~\bibnamefont
  {Boixo}}, \bibinfo {author} {\bibfnamefont {S.~V.}\ \bibnamefont {Isakov}},
  \bibinfo {author} {\bibfnamefont {V.~N.}\ \bibnamefont {Smelyanskiy}},
  \bibinfo {author} {\bibfnamefont {R.}~\bibnamefont {Babbush}}, \bibinfo
  {author} {\bibfnamefont {N.}~\bibnamefont {Ding}}, \bibinfo {author}
  {\bibfnamefont {Z.}~\bibnamefont {Jiang}}, \bibinfo {author} {\bibfnamefont
  {M.~J.}\ \bibnamefont {Bremner}}, \bibinfo {author} {\bibfnamefont {J.~M.}\
  \bibnamefont {Martinis}},\ and\ \bibinfo {author} {\bibfnamefont
  {H.}~\bibnamefont {Neven}},\ }\bibfield  {title} {\bibinfo {title}
  {Characterizing quantum supremacy in near-term devices},\ }\href
  {https://www.nature.com/articles/s41567-018-0124-x} {\bibfield  {journal}
  {\bibinfo  {journal} {Nature Physics}\ }\textbf {\bibinfo {volume} {14}},\
  \bibinfo {pages} {595} (\bibinfo {year} {2018})}\BibitemShut {NoStop}%
\bibitem [{\citenamefont {Pan}\ \emph {et~al.}(2020)\citenamefont {Pan},
  \citenamefont {Zhou}, \citenamefont {Li},\ and\ \citenamefont
  {Zhang}}]{pan2020contracting}%
  \BibitemOpen
  \bibfield  {author} {\bibinfo {author} {\bibfnamefont {F.}~\bibnamefont
  {Pan}}, \bibinfo {author} {\bibfnamefont {P.}~\bibnamefont {Zhou}}, \bibinfo
  {author} {\bibfnamefont {S.}~\bibnamefont {Li}},\ and\ \bibinfo {author}
  {\bibfnamefont {P.}~\bibnamefont {Zhang}},\ }\bibfield  {title} {\bibinfo
  {title} {Contracting arbitrary tensor networks: general approximate algorithm
  and applications in graphical models and quantum circuit simulations},\
  }\href {https://journals.aps.org/prl/abstract/10.1103/PhysRevLett.125.060503}
  {\bibfield  {journal} {\bibinfo  {journal} {Physical Review Letters}\
  }\textbf {\bibinfo {volume} {125}},\ \bibinfo {pages} {060503} (\bibinfo
  {year} {2020})}\BibitemShut {NoStop}%
\bibitem [{\citenamefont {Pan}\ and\ \citenamefont
  {Zhang}(2022)}]{pan2022simulation}%
  \BibitemOpen
  \bibfield  {author} {\bibinfo {author} {\bibfnamefont {F.}~\bibnamefont
  {Pan}}\ and\ \bibinfo {author} {\bibfnamefont {P.}~\bibnamefont {Zhang}},\
  }\bibfield  {title} {\bibinfo {title} {Simulation of quantum circuits using
  the big-batch tensor network method},\ }\href
  {https://journals.aps.org/prl/abstract/10.1103/PhysRevLett.128.030501}
  {\bibfield  {journal} {\bibinfo  {journal} {Physical Review Letters}\
  }\textbf {\bibinfo {volume} {128}},\ \bibinfo {pages} {030501} (\bibinfo
  {year} {2022})}\BibitemShut {NoStop}%
\bibitem [{cod()}]{code}%
  \BibitemOpen
  \href@noop {} {}\bibinfo {howpublished} {\url
  {https://github.com/bnuliujing/tn-for-unsup-ml}}\BibitemShut {NoStop}%
\end{thebibliography}%

\clearpage
\onecolumngrid
\appendix

\section{Autoregressive Matrix Product States}
\subsection{Model description}
Here we give a more detailed description of the Autoregressive Matrix Product States (AMPS) proposed in the main text. To fully exploit the representation power with the product of MPSs, we decompose the joint probability using the chain rule of probabilities in Eq.~(3) in the main text, which has recently been used in variational autoregressive networks (VAN)~\cite{wu2019solving} for solving statistical mechanics problem.
This decomposition enjoys tractable normalization since the normalization factor $\widetilde{Z}$ is exactly equal to 1:
\begin{align}
    \widetilde{Z}
    &=\sum_{x_1 x_2 \cdots x_n} \Psi^{(1)}_{x_1}\Psi^{(2)}_{x_1 x_2}\cdots\Psi^{(n)}_{x_1 x_2 \cdots x_n} \nonumber \\
    &=\sum_{x_1x_2 \cdots x_{n-1} }\Psi^{(1)}_{x_1}\Psi^{(2)}_{x_1 x_2}\cdots\Psi^{(n-1)}_{x_1 x_2 \cdots x_{n-1}}\sum_{x_{n}}\Psi^{(n)}_{x_1 x_2 \cdots x_n} \nonumber \\
    &=\sum_{x_1x_2 \cdots x_{n-2} }\Psi^{(1)}_{x_1}\Psi^{(2)}_{x_1 x_2}\cdots\Psi^{(n-2)}_{x_1 x_2 \cdots x_{n-2}}\sum_{x_{n-1}}\Psi^{(n-1)}_{x_1 x_2 \cdots x_{n-1}} \nonumber \\
    &=\cdots \nonumber \\
    &=\sum_{x_1 }\Psi^{(1)}_{x_1} = 1.
\end{align}
The above equations hold because each factor $\Psi^{(i)}_{x_1 x_2 \cdots x_i}$ is designed to be normalized: $\sum_{x_i}\Psi^{(i)}_{x_1 x_2 \cdots x_i}=1$.

As a concrete example, consider $4$ binary variables $\{x_1, x_2, x_3, x_4\}$. The joint distribution is written as 
\begin{align}
    P(x_1,x_2,x_3,x_4)&=P(x_1)P(x_2 \vert x_1)P(x_3 \vert x_1,x_2)P(x_4 \vert x_1,x_2,x_3)\nonumber\\    &=\Psi^{(1)}_{x_1}\Psi^{(2)}_{x_1,x_2}\Psi^{(3)}_{x_1,x_2,x_3}\Psi^{(4)}_{x_1,x_2,x_3,x_4}
\end{align}
In particular, $\Psi^{(4)}_{x_1,x_2,x_3,x_4}$ expresses the conditional probability $P(x_4\vert x_1,x_2,x_3)$. In order to express such a normalized conditional probability, we first construct an MPS
\begin{equation}
    \label{sm-eq:mps}
    \psi_{x_1,x_2,x_3,x_4}=\sum_{\{\alpha_0,\alpha_1,\alpha_2,\alpha_3,\alpha_4\}}A^{(1)}_{x_1 \alpha_0 \alpha_1} A^{(2)}_{x_2 \alpha_1 \alpha_2} A^{(3)}_{x_3 \alpha_2 \alpha_3} A^{(4)}_{x_4 \alpha_3 \alpha_4},
\end{equation}
where $A^{(i)}_{x_i \alpha_{i-1} \alpha_i}\in\mathbb R^{d\times D_{i-1}\times D_{i}}$ is a rank-3 tensor and $D_0=D_4=1$. 

\begin{figure}[!htbp]
    \centering
    \includegraphics[width=1\columnwidth]{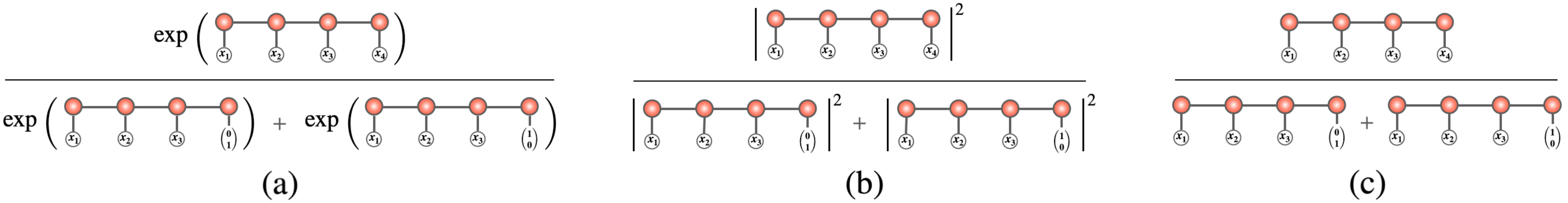}
    \caption{Different ways in representing the conditional probabilities by normalizing the MPS.}
    \label{fig:func}
\end{figure}

In the main text, we have introduced three different ways to ensure that the conditional probability $\Psi^{(i)}_{x_1 x_2 \cdots x_i}$ is normalized: using non-negative rank-3 tensors, taking an exponential function, called \textit{AMPS (Exponential)}, and adopting a squared norm as in the Born machine~\cite{han2018unsupervised}, called \textit{AMPS (Square norm)}:
\begin{align}
    \Psi^{(4)}_{x_1,x_2,x_3,x_4}&=\frac{\psi_{x_1,x_2,x_3,x_4}}{\sum_{x_4}\psi_{x_1,x_2,x_3,x_4}} \label{sm-eq:non-negative} \\
    \Psi^{(4)}_{x_1,x_2,x_3,x_4}&=\frac{\exp{(\psi_{x_1,x_2,x_3,x_4})}}{\sum_{x_4}\exp{(\psi_{x_1,x_2,x_3,x_4})}} \label{sm-eq:exp} \\
    \Psi^{(4)}_{x_1,x_2,x_3,x_4}&=\frac{\vert \psi_{x_1,x_2,x_3,x_4}\vert^2}{\sum_{x_4}\vert \psi_{x_1,x_2,x_3,x_4}\vert ^2},\label{sm-eq:square}
\end{align}
and the schematic representations are shown in Fig.~\ref{fig:func}. The main difficulty in training the AMPS model is that the contraction of tensors without using canonicalization usually causes numerical issues: the contraction may result in numerical underflow or overflow. In our implementations, we initialize each rank-3 tensor $A^{(i)}_{x_i \alpha_{i-1} \alpha_i}$ in such a way that $A^{(i)}_{x_i \alpha_{i-1} \alpha_i}$ with $x_i$ index fixed is equal to $I+\hat{A}$ where $I$ is the identity matrix and $\hat{A}$ is a learnable perturbation matrix which is initialized with small values.
If every learnable rank-3 tensor is non-negative, we term it \textit{AMPS (Non-negative I)}. Another way to ensure numerical stability is that in the contraction process, i.e. vector-matrix multiplication from the leftmost to the rightmost, the vector is scaled by dividing the maximum value. In the case of non-negative parameterization, we term it \textit{AMPS (Non-negative \textrm{II})}.

\begin{figure}[!htbp]
    \centering
    \includegraphics[width=0.5\columnwidth]{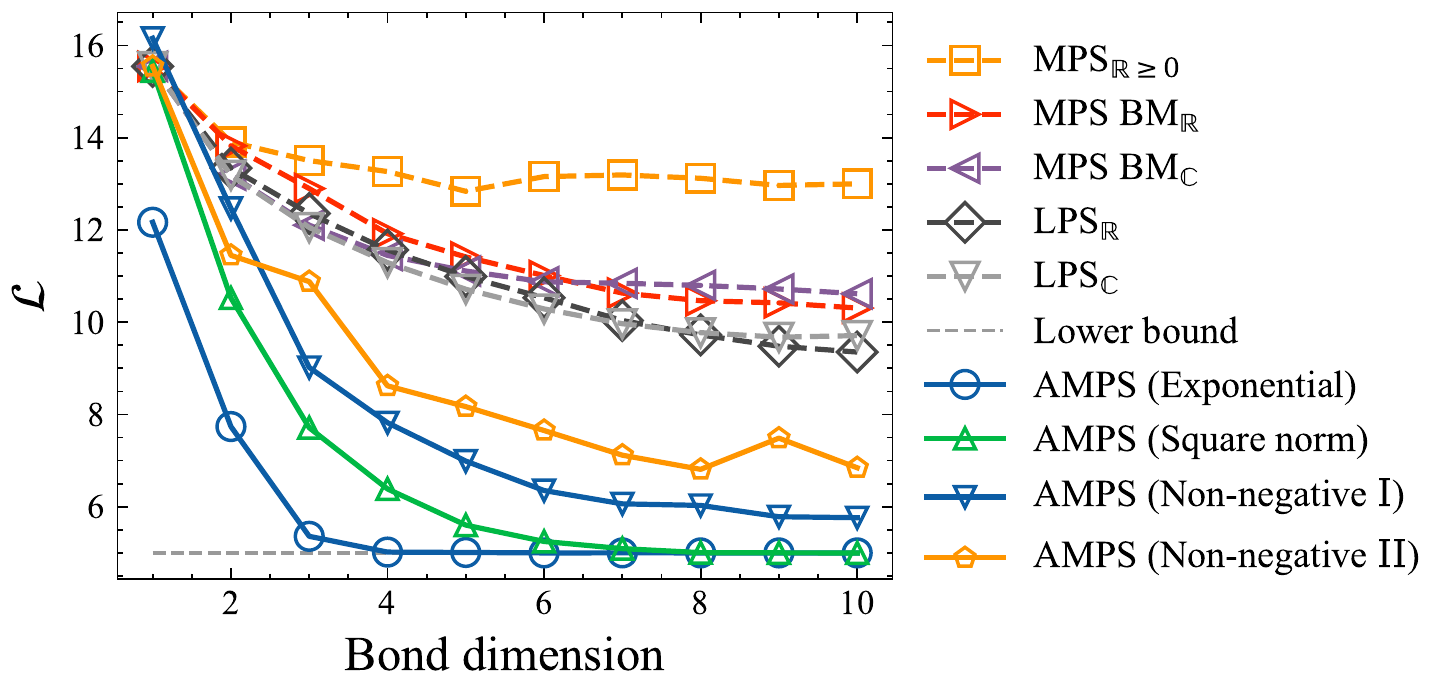}
    \caption{For the real-world dataset Lymphography shown in the main text, we provide extra results of AMPS (Square norm), AMPS (Non-negative I), and AMPS (Non-negative II).}
    \label{fig:compare_amps_1}
\end{figure}

\begin{figure}[!htbp]
    \centering
    \includegraphics[width=\columnwidth]{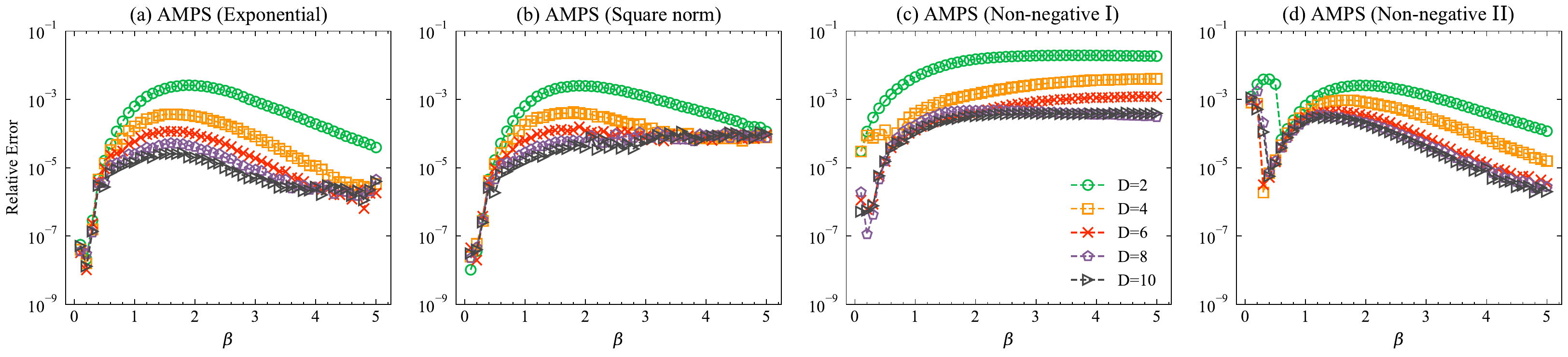}
    \caption{For a SK model with 20 spins, we show the relative error of free energy obtained using (a) AMPS (Exponential), (b) AMPS (Square norm), (c) AMPS (Non-negative \textrm{I}) and (d) AMPS (Non-negative \textrm{II}).}
    \label{fig:compare_amps_2}
\end{figure}

We have investigated the influences of different parameterizations on the performances of the AMPS. For the real-world dataset Lymphograhy studied in the main text, we provide the additional results of AMPS (Square norm), AMPS (Non-negative I), and AMPS (Non-negative II) in Fig.~\ref{fig:compare_amps_1}. Compared to AMPS (Exponential) which achieves the lower bound with only a bond dimension larger than 3, we found that the AMPS (Square norm) needs a bond dimension larger than 6 to achieve the same lower bound of the NLL. As for the two non-negative AMPS models, neither of them has achieved the lower bound with a bond dimension larger than 10. Nevertheless, both of them have outperformed other tensor network models such as MPS Born machines and locally purified states (LPS). For the experiment of the SK model, as shown in Fig.~\ref{fig:compare_amps_2}, the AMPS (Exponential) gives the most accurate estimation of free energy compared to other AMPS models. We notice that for AMPS (Exponential), AMPS (Square norm), and AMPS (Non-negative I) in the low-temperature (large $\beta$) region, the relative error no longer decreases as expected, especially when we further increase the bond dimension. We guess that the perturbation matrix condition is violated and causes some numerical instability during the learning process. For the AMPS (Non-negative II) model, we find that the optimization is extremely difficult at a high  temperature (small $\beta$) region, resulting in a much larger relative error. However, it works pretty well as the temperature anneals down, showing a competitive performance compared to the AMPS (Exponential) model.

\subsection{Parameter sharing scheme}
In order to reduce the number of parameters needed to describe the joint distribution, we introduce a parameter sharing scheme, which is widely used in state-of-the-art neural network architecture such as Convolutional Neural Networks (CNN). A simple parameter sharing scheme for the AMPS model is to force all MPSs to share the whole parameters. For example, the conditional probability $P(x_3\vert x_1,x_2)$ is constructed by the MPS
\begin{equation}
    \psi_{x_1,x_2,x_3}=\sum_{\{\alpha_0,\alpha_1,\alpha_2,\alpha_3\}}A^{(1)}_{x_1 \alpha_0 \alpha_1} A^{(2)}_{x_2 \alpha_1 \alpha_2} \tilde{A}^{(3)}_{x_3 \alpha_2 \alpha_3}
\end{equation}
where $A^{(1)}_{x_1 \alpha_0 \alpha_1}$ and $A^{(2)}_{x_2 \alpha_1 \alpha_2}$ are the same rank-3 tensors used in Eq.~(\ref{sm-eq:mps}). The last tensor $\tilde{A}^{(3)}_{x_3 \alpha_2 \alpha_3}=A^{(3)}_{x_3 \alpha_2 \alpha_3}[::0]$ is the first frontal slice of $A^{(3)}_{x_3 \alpha_2 \alpha_3}$ in Eq.~(\ref{sm-eq:mps}), which is a $2\times D_2\times 1$ tensor. We call this model \textit{Shared-AMPS}. We also notice that this parameter sharing scheme is very similar to that used in Neural Autoregressive Distribution Estimator (NADE)~\cite{larochelle2011neural}. 

In practice, we find that this parameter sharing scheme is useful to improve the model generalization ability and greatly reduces the number of parameters. For Shared-AMPS, since every conditional probability is parameterized by the same set of rank-3 tensors, the number of parameters only scales $n D^2$.

\subsection{Deep-AMPS}
\begin{figure}[!htbp]
    \centering
    \includegraphics[width=0.25\columnwidth]{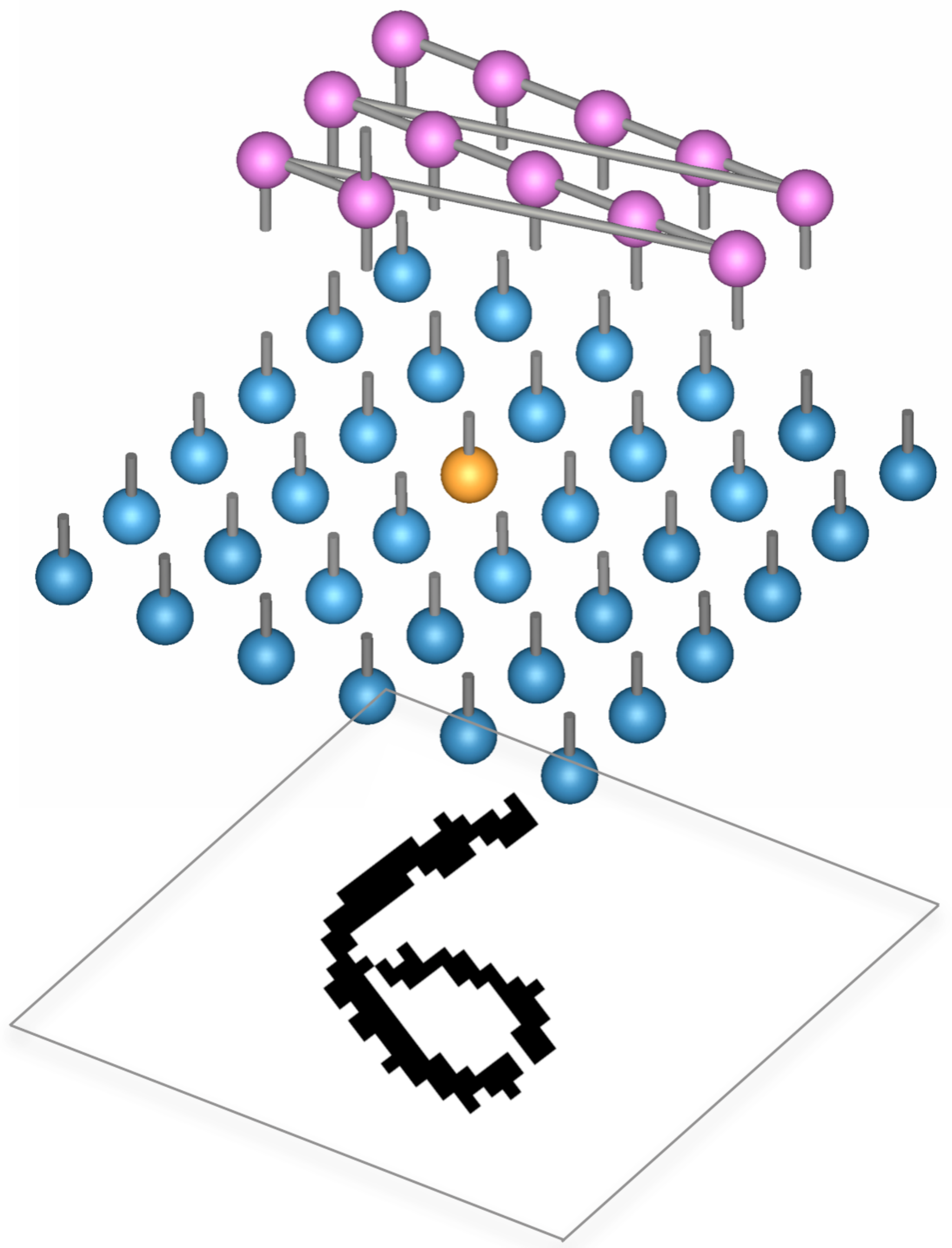}
    \caption{The first layer of Deep-AMPS.}
    \label{fig:deep_mps}
\end{figure}

For two-dimensional structures such as natural images, the pixel convolutional neural networks (PixelCNN)~\cite{van2016pixel} model is proposed to generate the images pixel by pixel using masked convolutional layers. Here we integrated the same idea into AMPS, by asking the MPSs to act on the spatial neighbors of a variable. We call this model \textit{Deep-AMPS}.
To put this differently, the CNN is a regularized version of general multi-layer perceptrons by considering the spatial dependencies in the image,
while the Deep-AMPS can be regarded as a regularized version of AMPS by considering the spatial dependencies in the two-dimensional data.
Fig.~\ref{fig:deep_mps} shows the first layer of the Deep-AMPS, where the yellow ball labels the target pixel. We first map the data into product states (blue balls), and then we use an MPS kernel (fuchsia balls) to contract with these product states. Similar to the A-type and B-type kernels in PixelCNN, the MPS also has two types. One is in the first layer, which does not contract with the target pixel and the dimension of each rank-3 tensor is $D\times2\times D$ where the physical bond dimension is 2. The other one is in the subsequent layers, which have one more tensor than the MPS in the first layer to contract with the vector in the target pixel position. Its dimension of each rank-3 tensor is $D\times D\times D$, except for the last tensor in the last convolutional layer's MPS, whose dimension is $D\times D\times2$ for outputting a binary distribution. Finally, we add a $\textit{Softmax}$ layer to obtain normalized conditional probability distributions.

\begin{figure}[!htbp]
    \centering
    \includegraphics[width=0.8\columnwidth]{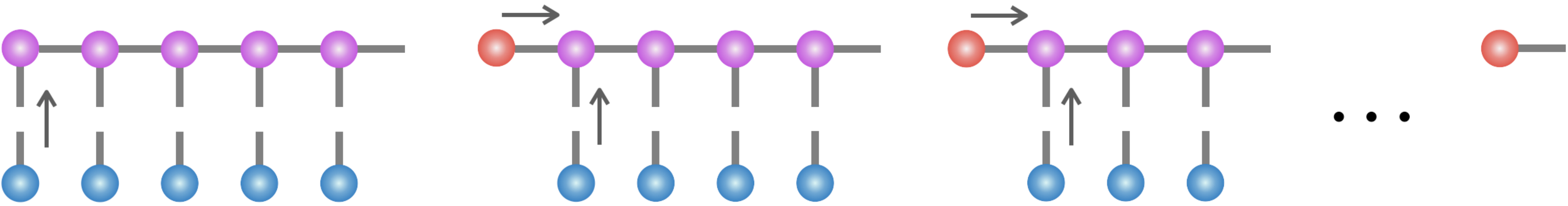}
    \caption{The algorithm of contracting the MPS and input vectors.}
    \label{fig:al_mps}
\end{figure}

The process of contracting the MPS in the first layer is displayed in Fig.~\ref{fig:al_mps}, where blue balls label product states and fuchsia balls label tensors in the MPS. After contracting the MPS and input vectors, an $D$-dimensional vector is produced, which is the input vector for the next layer. The computation complexity of this algorithm in a general layer (which means the dimension of a rank-3 tensor in the MPS is $D^3$) is $\mathcal{O}(D^3)$, where $D$ is the bond dimension of the MPS.

\subsection{Two-dimensional tensor network representation of RBM and AMPS}
For completeness, we first give a compact description of how to construct the two-dimensional tensor network representation of the RBM as proposed in~\cite{li2021boltzmann}.
In the RBM, the joint distribution of $n$ visible variables and $m$ hidden variables are defined as
\begin{align}
P(\x) =\frac{1}{Z}\sum_{\mathbf h}e^{J_{ij}x_ih_j+\theta_ix_i+\theta_jh_j}
=\frac{1}{Z}\prod_{i=1}^ne^{x_i\theta_i}\prod_{j=1}^m\Phi^{(j)}_{x_1,x_2,...,x_n},
\end{align}
where $J_{ij}$ denotes couplings between visible variable $i$ and hidden variable $j$; $\theta_i$ and $\theta_j$ denote external field of visible node $i$ and hidden node $j$ respectively, and
\begin{equation}
\Phi^{(j)}_{x_1,x_2,...,x_n}=2\cosh\left(\sum_{i=1}^nJ_{ij}x_i+\theta_j\right).
\end{equation}
That is, the joint distribution $P(\x)$ is factorized to the product of factors $\Phi$, each of which accepts a configuration of the visible variable $\x$ as input, and outputs a scalar value. When the visible variables are discrete, each $\Phi$ can be treated as a tensor $\mathcal {A}_j$ in the linear space with dimension $2^n$. Furthermore, $\mathcal{A}_j$ is a sum over two rank-one tensors $\prod_{i=1}^n \exp(J_{ij}x_i+\theta_j)$ and $\prod_{i=1}^n \exp(-J_{ij}x_i-\theta_j)$, hence is in a Canonical Polyadic (CP) form~\cite{MAL-059} with CP rank equals $2$. It is well known that a CP tensor can be converted exactly to an MPS with a bond dimension equal to its CP rank~\cite{oseledets2011tensor}. In this sense, an RBM can be regarded as a Correlator Product States~\cite{glasser2018neural,clark2018unifying}. Further, we notice that each visible variable $i$ can be regarded as a copy tensor with dimension $m+1$, and a copy tensor can be also converted to a MPS $\mathcal B_i$ with bond dimension $2$. Connecting every pair of $\mathcal B_i$ and $\mathcal {A}_j$ by contracting the associated couplings matrices between them, we arrive at a two-dimension tensor network representation of RBM as illustrated in Fig.~1(c) in the main text, where each row corresponds to an MPS with $\mathcal A$, and each column corresponds to an MPS with $\mathbf{B}$, all have bond dimension equals $2$. For more details about the two-dimensional representation, we refer to~\cite{li2021boltzmann}.

\begin{figure}[!htbp]
    \centering
    \includegraphics[width=\columnwidth]{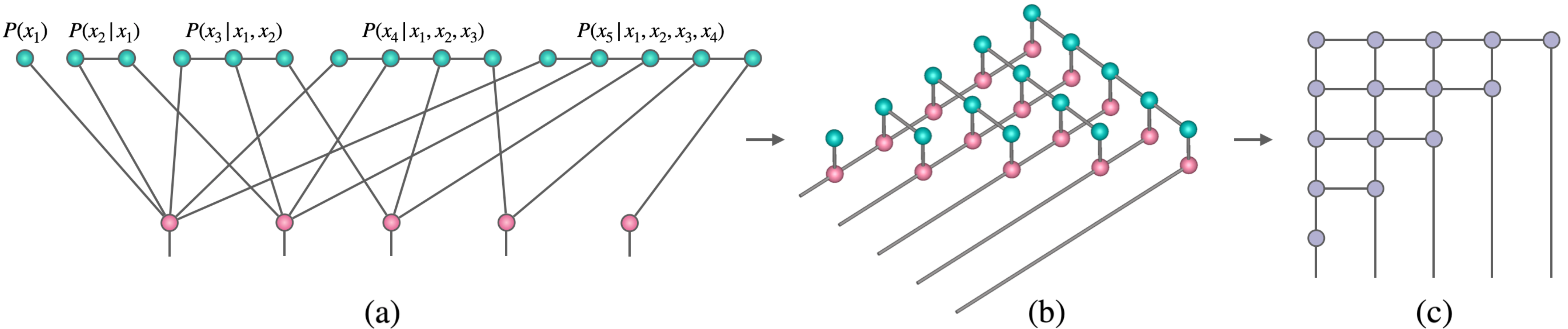}
    \caption{Illustration of converting the AMPS with a linear normalization function to the two-dimensional tensor network representation. (a) The graphical representation of AMPS with five binary variables. (b) After converting the copy tensors to MPSs, we get the three-dimensional structure. (c) The two-dimensional tensor network representation of AMPS.}
    \label{fig:2dtn}
\end{figure}

Analogous to the construction of the two-dimensional tensor network representation of RBM, in Fig.~\ref{fig:2dtn} we illustrate the two-dimensional tensor network representation for AMPS when $\Psi$ is a linear function of MPS $\psi$.
First, the joint probability distribution of five binary variables $\{x_1,x_2,x_3,x_4,x_5\}$ are converted to a tensor network in Fig.~\ref{fig:2dtn}(a), where the green MPSs are $\{\psi\}$ in Eq.~(\ref{sm-eq:mps}), and the red circles are copy tensors labeling variables. Here the conditional probability $\Psi$ is a linear function of $\psi$, i.e. normalized by the $l_1$ norm of $\psi$ with $\psi$ non-negative. Then we convert the copy tensors (red circles) to MPSs. For example, for variable $x_1$, the copy tensor with dimension 5 is converted to an MPS of length 5, as $\raisebox{-1ex}{\includegraphics[scale=0.25, trim={0cm 0.cm 0.cm 0cm}, clip]{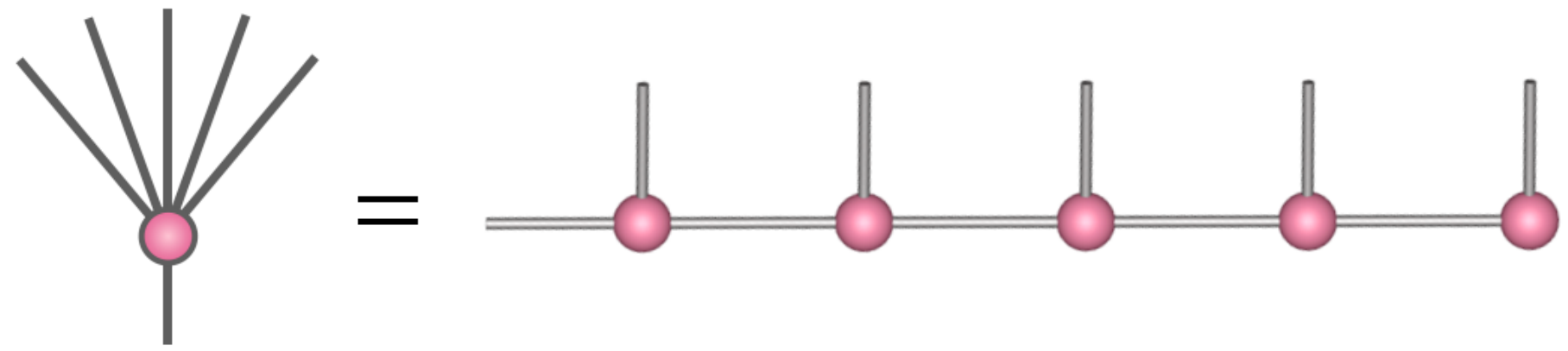}}$. By re-arranging the position of the MPSs, we arrive at Fig.~\ref{fig:2dtn}(b). Finally, by contracting the common indices between the green MPSs ($\Psi$) and the red MPSs (variables), we obtain the two-dimensional tensor network representation of AMPS, shown in Fig.~\ref{fig:2dtn}(c). 
If the $\Psi$ is a non-linear function of MPS $\psi$, then the tensor network representation would be more complicated, saturating a two-dimensional structure, as the representation in Fig.~1(d) of the main text denoted with solid boxes.

\subsection{Computational complexity}
In Table~\ref{tab:complexity} we list the number of parameters, training, and sampling complexity for various models.
For AMPS and Shared-AMPS, the training and sampling computational cost is $\mathcal{O}(n^2 D^2)$ and $\mathcal{O}(n D^2)$ respectively. For Deep-AMPS, the training computation cost is $\mathcal{O}(n D^3 l)$ per layer, and the sampling computation cost is $\mathcal{O}(n^2 D^3 l)$ where $l$ is the length of MPS. For an RBM with $n$ visible and $m$ hidden neurons, the training and sampling computational cost is $\mathcal{O}(knm)$ where $k$ is the number of Contrastive Divergence (CD) steps.

We also notice that in the sampling process of autoregressive models, the sampled pixel needs to be given as the input back into the model. This usually requires $n$ repeated computation for MADE, PixelCNN as well as our Deep-AMPS. However, for both AMPS and Shared-AMPS since the sampled pixel has already been used for contracting the next MPS tensor, which is equivalent to the contracting process during training, redundant computations are not necessary. In our experiment, AMPS, MADE PixelCNN takes 0.22s, 1.12s, and 1.32s to sample one image of binarized MNIST image on Nvidia Tesla V100 respectively.

\begin{table}[!htbp]
\begin{tabular}{@{}llll@{}}
\toprule
Model                 & \# parameters          & Training complexity    & Sampling complexity      \\ \midrule
AMPS                  & $\mathcal{O}(n^2 D^2)$ & $\mathcal{O}(n^2 D^2)$ & $\mathcal{O}(n^2 D^2)$   \\
Shared-AMPS           & $\mathcal{O}(n D^2)$   & $\mathcal{O}(n D^2)$   & $\mathcal{O}(n D^2)$     \\
Deep-AMPS (per layer) &   $\mathcal{O}(D^3 l)$ & $\mathcal{O}(n D^3 l)$ & $\mathcal{O}(n^2 D^3 l)$ \\
MADE (per layer)      & $\mathcal{O}(n^2)$     & $\mathcal{O}(n^2)$     & $\mathcal{O}(n^3)$       \\
RBM                   & $\mathcal{O}(nm)$      & $\mathcal{O}(knm)$     & $\mathcal{O}(knm)$       \\ \bottomrule
\end{tabular}
\caption{Number of parameters and time complexity. Here $n$ is the input size, $D$ is the bond dimension, $l$ is the length of MPS used in Deep-AMPS, $m$ is the hidden neurons in RBM, and $k$ is the CD steps in the training of RBM.}
\label{tab:complexity}
\end{table}

\section{AMPS for generative modeling}
\subsection{A gentle introduction to generative modeling}
We first give a gentle introduction to the task of generative modeling. In generative modeling, we try to train a model to represent the target data distribution. Given empirical data distribution $P_{\text{data}}(\x)$ and model distribution $P(\x;\btheta)$ parameterized by $\btheta$, the Kullback-Leibler (KL) divergence measures the closeness of two probability distribution:
\begin{align}
    \KLD{P_{\text{data}}(\x)}{P(\x;\btheta)}&=\sum_{\x}P_{\text{data}}(\x)\ln \left(\frac{P_{\text{data}}(\x)}{P(\x;\btheta)}\right) \nonumber \\
    &=\sum_{\x}P_{\text{data}}(\x)\ln P_{\text{data}}(\x) - \sum_{\x}P_{\text{data}}(\x)\ln P(\x;\btheta). 
\end{align}
The first term on the RHS is the entropy of the empirical data distribution, which is equal to $-\ln \vert \mathcal{T}\vert$ if there is no duplication in the dataset. 
By replacing $P_{\text{data}}(\x)$ with $\frac{1}{\vert\mathcal{T}\vert}\sum_{\x^{\prime}\in\mathcal{T}}\delta(\x-\x^{\prime})$, the second term is the negative log-likelihood (NLL):
\begin{align}\label{sm-eq:nll}
    \mathcal{L}&=- \sum_{\x}P_{\text{data}}(\x)\ln P(\x;\btheta) \nonumber \\
    &= -\frac{1}{\vert\mathcal{T}\vert}\sum_{\x\in\mathcal{T}}\ln P(\x;\btheta).
\end{align}
Since the KL divergence is non-negative, we have $\mathcal{L}\geq -\sum_{\x}P_{\text{data}}(\x)\ln P_{\text{data}}(\x)$, which means that the entropy of the empirical data distribution provides the lower bound for the NLL. When $\mathcal{L}=-\sum_{\x}P_{\text{data}}(\x)\ln P_{\text{data}}(\x)$, the model distribution is exactly the empirical data distribution. The pseudocode of \textit{AMPS for generative modeling} is presented in the Algorithm~\ref{algo-gm}.

\begin{algorithm}[H]
	\caption{\textit{AMPS for generative modeling}\label{algo-gm}}
	\begin{algorithmic}  
		\Require
		MPS parameters $\btheta$, dataset $\mathcal{T} =\{\x_1, \x_2, \cdots, \x_{\vert \mathcal{T} \vert}\}$, batch size $N_{\text{bs}}$, learning rate $\eta$
		\Ensure
		Well trained model $P(\x;\btheta)$, NLL $\mathcal{L}$.
		\State Initialize MPS parameters $\btheta$
		\While{Stopping criterion is not satisfied}
		\State Sample a batch of $N_{\text{bs}}$ data points from dataset $\mathcal{T}$.
		\State Compute NLL $\mathcal{L}_{\btheta}$ using Eq.~(\ref{sm-eq:nll})
		\State Compute the gradients $\nabla_{\btheta}\mathcal{L}_{\btheta}$ 
		\State Update MPS parameters using the gradients and an optimizer (such as Adam~\cite{kingma2014adam}). 
	    \EndWhile
		\If{Sample from the learned distribution}
		\For{$i=1, 2, \cdots, n$}
		\State Compute $P(x_i \vert x_1, \cdots, x_{i-1})$
		\State Sample $x_i$ from $P(x_i \vert x_1, \cdots, x_{i-1})$
		\EndFor
		\EndIf
		\State  Return NLL $\mathcal{L}$ and new samples from the learned distribution.
	\end{algorithmic} 
\end{algorithm}

\subsection{More results on random datasets}
\begin{figure}[!htbp]
    \centering
    \includegraphics[width=\linewidth]{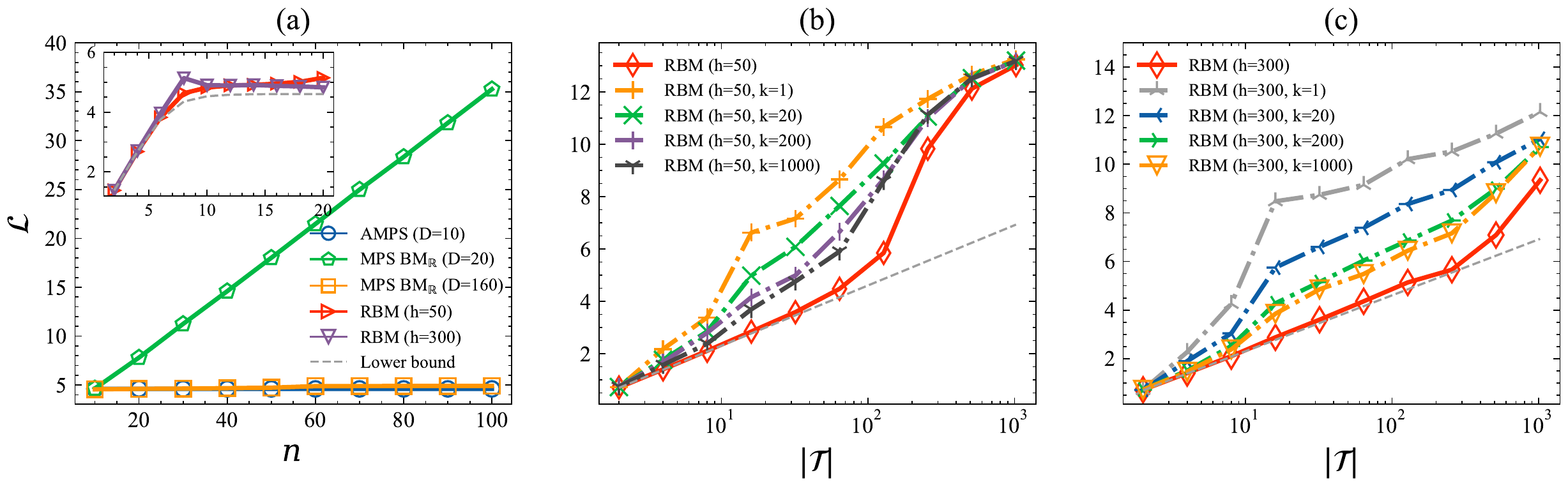}
    \caption{(a) For random dataset with fixed sample size $\vert \mathcal{T} \vert=100$, we plot the NLL versus the system size $n$ for different models. (b) (c) For random dataset with 20 variables, we show the performance of the RBM with hidden variables $h=50$ and $h=300$.}
    \label{fig:rbm-CD}
\end{figure}

For random samples drawn from the Bernoulli distribution, since we are only interested in expressive power, we try to obtain an NLL that is as small as possible. In all the experiments, we have used batch size equal to the number of samples $\vert \mathcal{T} \vert$, learning rate $0.001$, and 10,000 training steps to ensure that the optimization converges.
In addition to the experiments with a fixed number of variables presented in the main text, we also conducted numerical experiments with a fixed number of samples $\vert \mathcal{T}\vert=100$, the performance is shown in Fig.~\ref{fig:rbm-CD}(a). Notice that for a small value of $n$, the total number of configurations $2^n$ is smaller than $\vert \mathcal{T}\vert=100$, so the lower bound for $\mathcal{L}$ should be calculated according to $-\sum_{\x}P_{\text{data}}(\x)\ln P_{\text{data}}(\x)$. We can see from the figure that with a larger bond dimension $D$, the MPS Born machine obtains smaller $\mathcal{L}$. When bond dimension $D=160$ which is larger than $\vert \mathcal{T}\vert$, the MPS Born machine gives $\mathcal{L}=\ln \vert \mathcal{T}\vert$. Remarkably, our AMPS model can achieve the same performance with a much smaller bond dimension $D=10$.

In the training process of RBM, the gradients can be exactly computed involving exhaustive enumerating of all configurations, or approximately calculated such as the Contrastive Divergence (CD) algorithm. Fig.~\ref{fig:rbm-CD}(b)(c) show the NLL on the training dataset obtained with different gradient-computing methods with hidden variables $h=50$ and $300$. Apparently, when the CD step increases, the model performs better and gives a closer to the exact gradients. However, none of these RBM models can reach the theoretical lower bound $\ln \vert \mathcal{T} \vert$.

\subsection{More results on real-world datasets}
To fully demonstrate the expressive power, we have also performed extensive experiments on various real-world datasets studied in~\cite{glasser2019expressive}: Biofam dataset from the Swiss Household Panel biographical survey~\cite{muller2007classification}, datasets from the UCI Machine Learning Repository~\cite{Dua:2019}: Lymphography~\cite{lymphodataset}, SPECT Heart, Congressional Voting Records, Primary Tumor~\cite{lymphodataset} and Solar Flare. The dataset statistics are summarized in Table~\ref{tab:realistic-data}.

\begin{table}[!htbp]
\centering
\begin{tabular}{@{}cccc@{}}
    \toprule
    Dataset      & $\vert \mathcal{T} \vert$ & $n$  & $d$ \\ \midrule
    Biofam       & 2000      & 16 & 8           \\
    Lymphography & 148       & 19 & 8           \\
    SPECT Heart        & 187       & 23 & 2           \\
    Congressional Voting Records        & 435       & 17 & 3           \\
    Primary Tumor        & 339       & 17 & 4           \\
    Solar Flare        & 1065      & 13 & 8           \\ \bottomrule
\end{tabular}
\caption{Statistics of real-world datasets. Here $\vert \mathcal{T} \vert$ denotes the total number of samples, $n$ denotes the number of variables and $d$ denotes the number of categories.}
\label{tab:realistic-data}
\end{table}

We compare against the following tensor network models:
\begin{itemize}
    \item Positive MPS ($\text{MPS}_{\mathbb{R}\geq 0}$): decomposing the joint distribution $P(\x)$ into an MPS with non-negative tensors $A^{(i)} \in \mathbb{R}_{\geq 0}^{d \times D_{i-1} \times D_i}$:
    \begin{align}
        P(\x)&=\frac{1}{Z} T(x_1,x_2,...,x_n),\\
        T(x_1,x_2,\cdots,x_n)&=\sum_{\{\alpha_i=1\}}^D
        A^{(1)}_{x_1\alpha_0 \alpha_1}
        A^{(2)}_{x_2\alpha_1 \alpha_2} \cdots
        A^{(n)}_{x_{n}\alpha_{n-1} \alpha_n}.
    \end{align}
    \item MPS Born machines ($\text{MPS BM}_{\mathbb{R}}$): the $P(\x)$ is modeled by squared norm of a quantum state $\Psi$ with $A^{(i)} \in \mathbb{R}^{d \times D_{i-1} \times D_i}$:
    \begin{align}
        P(\x)&=\frac{1}{Z}\left |\Psi(x_1,x_2,...,x_n)\right |^2,\\
        \Psi(x_1,x_2,...,x_n)&=\sum_{\{\alpha_i=1\}}^D
        A^{(1)}_{x_1\alpha_0 \alpha_1}
        A^{(2)}_{x_2\alpha_1 \alpha_2} \cdots
        A^{(n)}_{x_{n}\alpha_{n-1} \alpha_n}.
    \end{align}
    \item  MPS Born machines with complex entries ($\text{MPS BM}_{\mathbb{C}}$): same as MPS Born machines but with complex tensors $A^{(i)} \in \mathbb{C}^{d \times D_{i-1}\times D_i}$.
    \item Locally Purified States ($\text{LPS}_{\mathbb{R}}$): proposed in~\cite{glasser2019expressive} for probabilistic modeling. 
    The Locally Purified States model the joint distribution by introducing additional \textit{purification dimension} in MPS tensors and the $P(\x)$ can be written as
    \begin{equation}
        P(\x)=\frac{1}{Z} T(x_1,x_2,\cdots,x_n),
    \end{equation}
    where
    \begin{align}
        T(x_1,x_2,\cdots,x_n)
        =\sum_{\{\alpha_i,\alpha'_i=1\}}^D
        \sum_{\beta_i=1}^M
        A^{(1)}_{x_1\alpha_0 \alpha_1 \beta_1}
        A^{(1)}_{x_1\alpha'_0 \alpha'_1 \beta_1}
        A^{(2)}_{x_2\alpha_1 \alpha_2 \beta_2}
        A^{(2)}_{x_2\alpha'_1 \alpha'_2 \beta_2}\cdots
        A^{(n)}_{x_{n}\alpha_{n-1} \alpha_n \beta_n}
        A^{(n)}_{x_{n}\alpha'_{n-1} \alpha'_n \beta_n}.
        \label{eq:LPS}
    \end{align}
    The $A^{(i)}\in\mathbb R^{2\times D_{i-1}\times D_{i}\times M_{i}}$ is a rank-4 tensor with $D_0 = D_n = 1$ and the indices $\beta_i$ denotes the purification dimension.
    \item Locally Purified State with complex entries ($\text{LPS}_{\mathbb{C}}$): same as $\text{LPS}_{\mathbb{R}}$ but with complex tensors $A^{(i)} \in \mathbb{C}^{d \times D_{i-1}\times D_i \times M_i}$, so the Eq.~(\ref{eq:LPS}) becomes:
    \begin{align}
        T(x_1,x_2,\cdots,x_n)
        =\sum_{\{\alpha_i,\alpha'_i=1\}}^D
        \sum_{\beta_i=1}^M
        A^{(1)}_{x_1\alpha_0 \alpha_1 \beta_1}
        \overline{A^{(1)}_{x_1\alpha'_0 \alpha'_1 \beta_1}}
        A^{(2)}_{x_2\alpha_1 \alpha_2 \beta_2}
        \overline{A^{(2)}_{x_2\alpha'_1 \alpha'_2 \beta_2}}\cdots
        A^{(n)}_{x_{n}\alpha_{n-1} \alpha_n \beta_n}
        \overline{A^{(n)}_{x_{n}\alpha'_{n-1} \alpha'_n \beta_n}},
    \end{align}
    where the $\overline{A^{(i)}}$ denotes the complex conjugate of $A^{(i)}$. 
\end{itemize}

In all experiments, we have used a batch size equal to the number of samples $\vert \mathcal{T}\vert$. For different models, the number of parameters as well as the learning rate and training steps are summarized in Table~\ref{tab:baseline}. The implementations of the reference tensor network methods as well as the datasets were downloaded from~\cite{download}.

\begin{table}[!htbp]
\centering
\begin{tabular}{@{}cccc@{}}
\toprule
Model                           & \# parameters   & learning rate & training steps \\ \midrule
AMPS                           & $dn(n+1)D^2/2$ & 0.001        & $10,000$       \\
Shared AMPS                    & $dnD^2$        & 0.001        & $10,000$       \\
$\text{MPS}_{\mathbb{R}\geq 0}$ & $dnD^2$        & $1$           & $10,000$       \\
$\text{MPS BM}_{\mathbb{R}}$        & $dnD^2$        & $1$           & $10,000$       \\
$\text{MPS BM}_{\mathbb{C}}$        & $2dnD^2$       & $1$           & $10,000$       \\
$\text{LPS}_{\mathbb{R}}$       & $2dnD^2$       & $10$          & $10,000$       \\
$\text{LPS}_{\mathbb{C}}$       & $4dnD^2$       & $10$          & $10,000$       \\ \bottomrule
\end{tabular}
\caption{Summary of the number of parameters, learning rate, and training steps for different models.}
\label{tab:baseline}
\end{table}

\begin{figure}[!htbp]
    \centering
    \includegraphics[width=\linewidth]{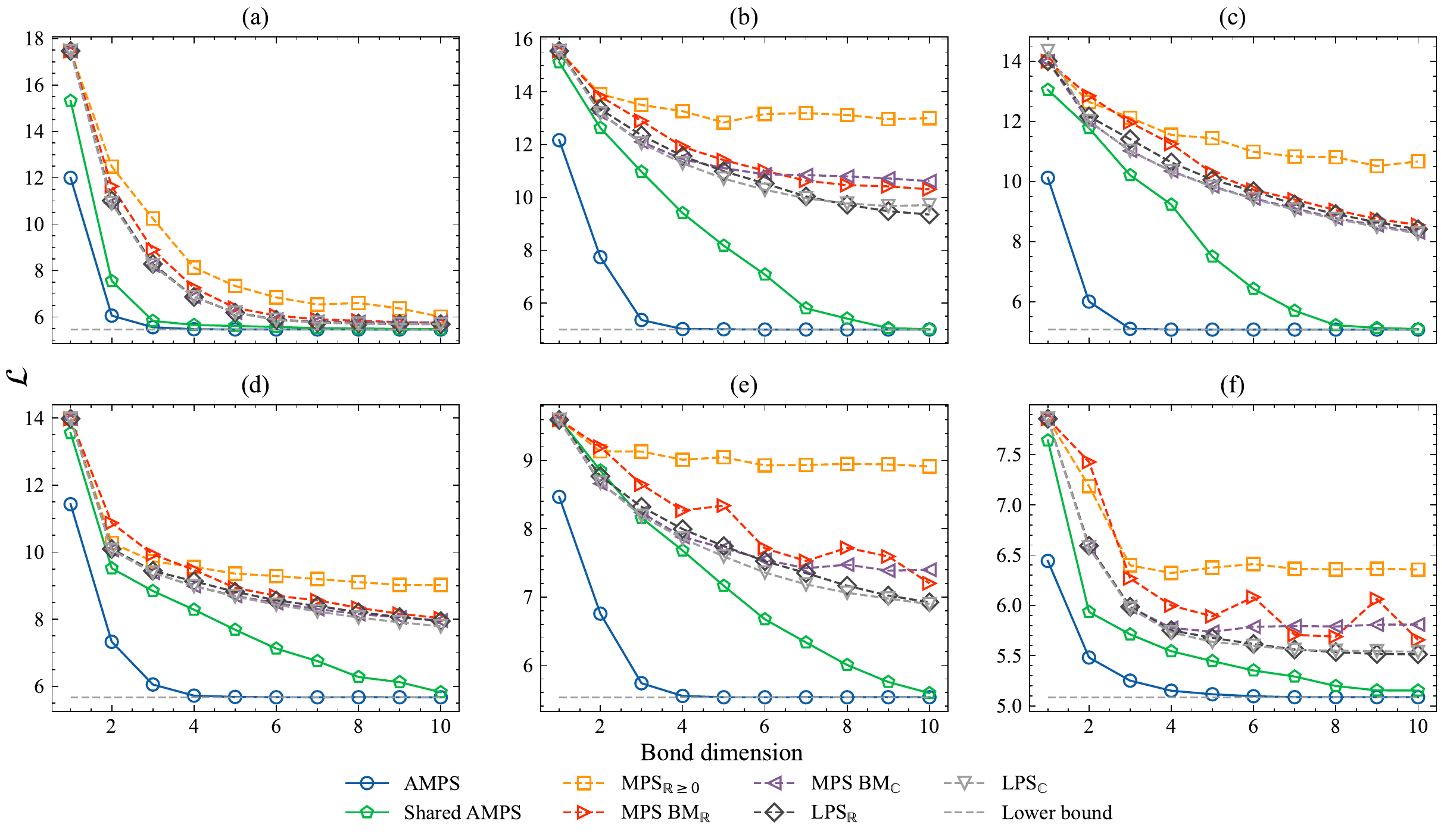}
    \caption{NLL as a function of bond dimension for different tensor network models on the real-world datasets: (a) Biofam data set: the family life states data from the Swiss Household Panel biographical survey~\cite{muller2007classification} and the datasets come from the UCI Machine Learning Repository~\cite{Dua:2019}: (b) Lymphography and  tumor domains~\cite{lymphodataset}, (c) SPECT Heart, (d) Congressional Voting Records, (e) Primary Tumor~\cite{lymphodataset}, (f) Solar Flare. The lower bounds for $\mathcal L$ are determined using the entropy of the empirical data distribution. All datasets are downloaded from~\cite{download}.}
    \label{fig:capacity-1}
\end{figure}

\begin{figure}[!htbp]
    \centering
    \includegraphics[width=\linewidth]{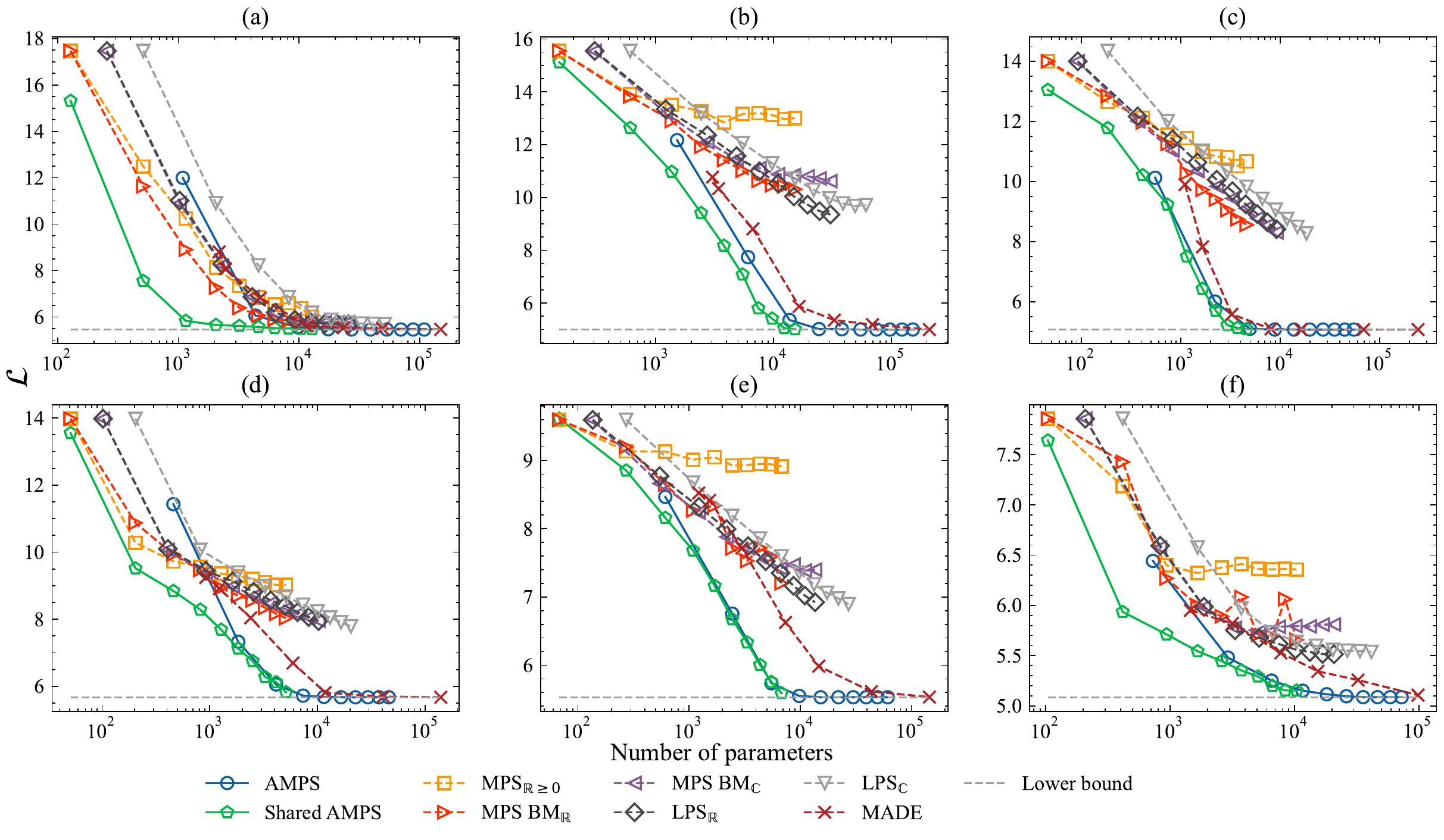}
    \caption{NLL as a function of the number of parameters for different tensor network models as well as neural network model MADE~\cite{germain2015made} on the real-world datasets.}
    \label{fig:capacity-2}
\end{figure}

We plot the NLL as a function of bond dimension $D$ for different tensor network models in Fig.~\ref{fig:capacity-1}. From the figure, we see that the AMPS model significantly outperforms other tensor network models by a large gap. Remarkably, even Shared-AMPS with fewer parameters is already significantly superior with the same bond dimension.
In Fig.~\ref{fig:capacity-2} we investigate how the number of parameters will influence the performances.
We also show the results given by MADE~\cite{germain2015made}, the state-of-the-art neural network autoregressive estimator. Surprisingly, the AMPS model can achieve much lower NLL compared to other models, including MADE, indicating that the AMPS model has much stronger expressive power than other models.

\subsection{More results on binarized MNIST dataset}
In the above experiments on random and real-world datasets, we only cared about the expressive power of the model while completely ignoring the generalization, i.e., evaluating the NLL on the test dataset and the ability to generate new samples. Here we investigate the generalization power of Shared-AMPS and Deep-AMPS on the binarized MNIST dataset, a standard dataset in generative modeling. After training the models on the MNIST dataset, we can sample from the learned distribution through the \textit{ancestral sampling}~\cite{han2018unsupervised,bishop2006pattern}. 
As a concrete example, consider again the case with four binary variables. To samples a new data point $\x$, we first determine $x_1$ from $\Psi_{x_1}^{(1)}$, then determine $x_2$ from $\Psi_{x_1,x_2}^{(2)}$ using $x_1$, then determine $x_3$ from $\Psi_{x_1,x_2,x_3}^{(3)}$ using $x_1$ and $x_2$, finally determine $x_4$ from $\Psi_{x_1,x_2,x_3,x_4}^{(4)}$ using $x_1$, $x_2$ and $x_3$.

\begin{figure}[!htbp]
    \centering
    \includegraphics[width=0.4\linewidth]{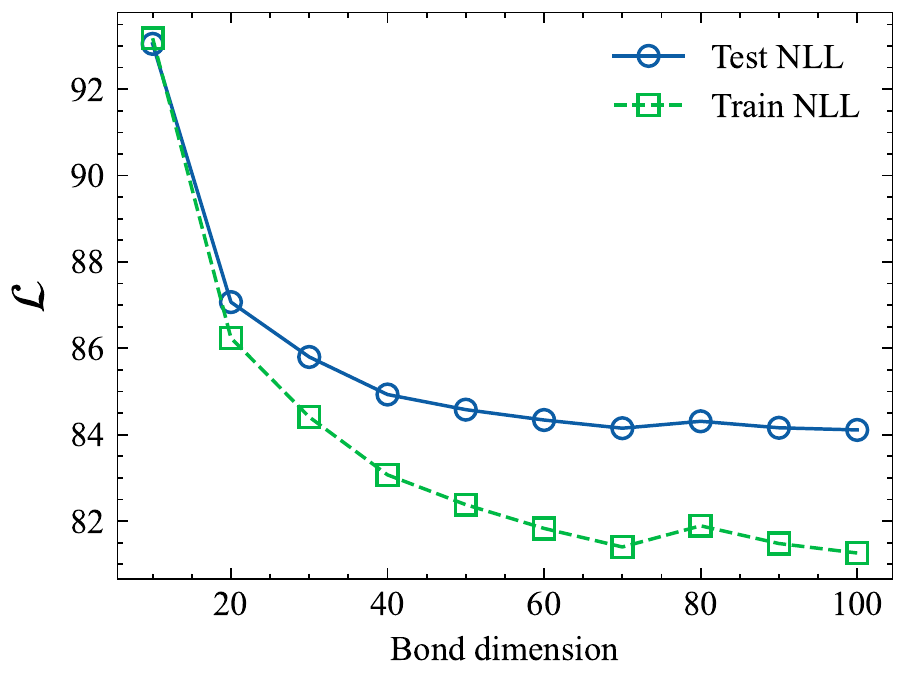}
    \caption{Training NLL and test NLL as a function of the bond dimension $D$ in the Shared-AMPS trained on the binary MNIST dataset using all $50,000$ training images.}
    \label{fig:nll-vs-D}
\end{figure}

For Shared-AMPS, we have used batch size 100, initial learning rate 0.1, epochs 50, and decaying the learning rate by 0.1 every 10 epochs. 
As for Deep-AMPS, we have used an MPS with a length of 24 (in the first layer), and bond dimension $D=14$. The network consists of 8 convolutional layers. We also use \textit{padding} to ensure that the input size of each layer is the same as the data size ($28\times28$).
In addition, in Fig.~\ref{fig:nll-vs-D} we plot the training NLL and test NLL as a function of bond dimension for the Shared-AMPS model. We find that as the bond dimension increases, the performance on the test dataset no longer improves, indicating an overfitting phenomenon with more parameters than we need on the dataset. The best test NLL is $84.1$ with the dimension $D=100$, as reported in the main text. To the best of our knowledge, this is the best result given by tensor network models. 

In our experiment, we compared the performance of AMPS with existing neural network models. Here we give a brief introduction to them:
\begin{enumerate}
    \item neural autoregressive distribution estimator (NADE)~\cite{larochelle2011neural}: an MLP (MultiLayer Perceptron)-based parameterization where each conditional probability $P(x_i | \x_{<i})$ is expressed as
    \begin{align}
        P(x_i | \x_{<i}) &= \textit{Softmax} \left(\boldsymbol{\alpha}_i \mathbf{h}_i + \mathbf{b}_i\right),\\
        \mathbf{h}_i &= \sigma\left( \mathbf{W}_{:, <i} \x_{<i} + \mathbf{c}\right),
    \end{align}
    where $\{\mathbf{W}, \mathbf{c}, \boldsymbol{\alpha}, \mathbf{b}\}$ are learnable parameters, $\sigma(\cdot)$ denotes the activation function, and the parameter $\mathbf{W}$ and $\mathbf{c}$ are shared in the calculation of different hidden vectors $\mathbf{h}_i$.
    \item masked autoencoder distribution estimator (MADE)~\cite{germain2015made}: an autoencoder-based parameterization where the connections (weight matrices) are masked to achieve the autoregressive property. For a simple 1-layer MADE, the conditional probability can be written as
    \begin{align}
        P(x_i | \x_{<i}) = \textit{Softmax}\left((\mathbf{M}_{:, i} \odot \mathbf{W}_{:, i}) \x + b_i \right),
    \end{align}
    where $\mathbf{M}$ is the masked matrix and $\{\mathbf{W}, \mathbf{b}\}$ are learnable parameters.
    \item pixel convolutional neural networks (PixelCNN)~\cite{van2016pixel}: a CNN-based parameterization where a masked convolution is adopted such that $x_i$ is only dependent on the variables left and above of $x_i$.
\end{enumerate}

\section{AMPS for reinforcement learning in statistical physics}
\subsection{A gentle introduction to reinforcement learning in statistical physics}
For reinforcement learning in statistical physics, we try to make the variational distribution $q(\s;\btheta)$ as close as possible to the equilibrium Boltzmann distribution $P(\s)=\exp{(-\beta E(\s))}/Z$. The closeness of two probability distributions is measured by the reverse KL divergence:
\begin{equation}
    \KLD{q(\s;\btheta)}{P(\s)}=\sum_{\s}q(\s;\btheta)\ln\left(\frac{q(\s;\btheta)}{P(\s)}\right)=\beta (F_q-F),
\end{equation}
where for the convention of statistical physics, we use $\s$ to denote the spin configuration $\s \in \{-1,+1\}^n$ and 
\begin{equation}
\label{sm-eq:free-energy}
    F_q=\frac{1}{\beta}\sum_{\s} q(\s;\btheta)\left[\beta E(\s)+\ln q(\s;\btheta)\right]
\end{equation}
is the variational free energy, and $F$ is the true free energy. 
In our numerical experiments, the variational distribution is parameterized by our AMPS model. For the Sherrington-Kirkpatrick (SK) model~\cite{sherrington1975solvable}, the energy $E(\s)$ is defined as 
\begin{equation}
\label{sm-eq:energy}
E(\s)=-\sum_{1\leq i<j\leq n}J_{ij}s_i s_j.
\end{equation}
By denoting $\hat{s}_i=P(s_i=+1)$, the entropy is given by
\begin{equation}
\label{sm-eq:entropy}
    \ln q(\s;\btheta)=\sum_{i=1}^{n}\left[s_i\ln\hat{s}_i+(1-s_i)\ln{(1-\hat{s}_i)} \right].
\end{equation}
We adopt the REINFORCE algorithm~\cite{williams1992simple} for estimating the gradient of the variational free energy with respect to model parameters:
\begin{equation}
\label{sm-eq:reinforce}
    \nabla_{\btheta}F_q=\frac{1}{\beta}\sum_{\s} q(\s;\btheta) \left[\beta E(\s)+\ln q(\s;\btheta)\right]\nabla_{\btheta} \ln q(\s;\btheta).
\end{equation}
The pseudocode of \textit{AMPS for reinforcement learning in statistical physics} is presented in the Algorithm~\ref{algo-rl}.

\begin{algorithm}[H]
	\caption{\textit{AMPS for reinforcement learning in statistical physics}\label{algo-rl} }
	
	\begin{algorithmic}  
		\Require
		MPS parameters $\btheta$, SK model parameters $\{J_{ij}\}$, batch size $N_{\text{bs}}$, learning rate $\eta$
		\Ensure
		Trained variational distribution $q(\s;\btheta)$, variational free energy $F_q$.
		\State Initialize MPS parameters $\btheta$
		\While{Stopping criterion is not satisfied}
		\State Generate a batch of $N_{\text{bs}}$ samples from the variational distribution $q(\s;\btheta)$ by the ancestral sampling
		\State Compute the energy and the entropy using Eq.~(\ref{sm-eq:energy}) and Eq.~(\ref{sm-eq:entropy}), then compute variational free energy $F_q$ using Eq.~(\ref{sm-eq:free-energy}).
		\State Compute the gradients $\nabla_{\btheta}F_q$ from Eq.~(\ref{sm-eq:reinforce})
		\State Update MPS parameters using the gradients and an optimizer (such as Adam~\cite{kingma2014adam}).
	    \EndWhile
	    \State Return Variational free energy $F_q$
	\end{algorithmic} 
\end{algorithm}

\subsection{More results on SK model}
\begin{figure}[!htbp]
    \centering
    \includegraphics[width=\linewidth]{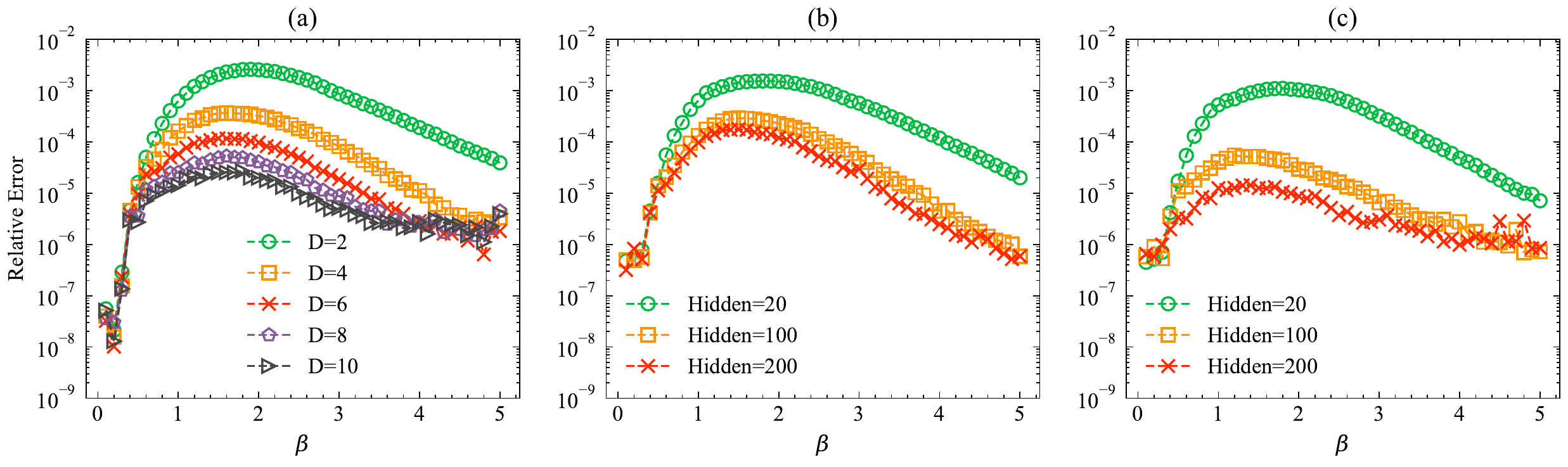}
    \caption{For the SK model with 20 spins, we plot the relative error of free energy for (a) AMPS with different bond dimensions, (b) a 2 layer VAN with different hidden neurons, and (c) a 3 layer VAN with different hidden neurons.}
    \label{fig:sk-vs-van}
\end{figure}

In our experiments, we follow the same experimental setup in~\cite{wu2019solving}. We have used a batch size of $10,000$ and a learning rate of 0.001. The main results are shown in the main text. In Table~\ref{tab:sk-hyper} we show the model details and the corresponding number of parameters. In Fig.~\ref{fig:sk-vs-van} we also show the relative error of free energy for AMPS with different bond dimensions as well as VAN with more hidden neurons or depths.

\begin{table}[!htbp]
\centering
\begin{tabular}{@{}ccc@{}}
\toprule
Model                        & Detail             & \# Parameters \\ \midrule
\multirow{5}{*}{AMPS}       & D=2                & 1600          \\
                             & D=4                & 6400          \\
                             & D=6                & 14400         \\
                             & D=8                & 25600         \\
                             & D=10               & 40000         \\ \midrule
\multirow{3}{*}{2 layer VAN} & 20 hidden neurons  & 800           \\
                             & 100 hidden neurons & 4000          \\
                             & 200 hidden neurons & 8000          \\ \midrule
\multirow{3}{*}{3 layer VAN} & 20 hidden neurons  & 1200          \\
                             & 100 hidden neurons & 14000         \\
                             & 200 hidden neurons & 48000         \\ \bottomrule
\end{tabular}
\caption{Model details for SK model with 20 spins.}
\label{tab:sk-hyper}
\end{table}

\subsection{Unbiased estimation by neural importance sampling}
\begin{figure}[!htbp]
    \centering
    \includegraphics[width=0.4\linewidth]{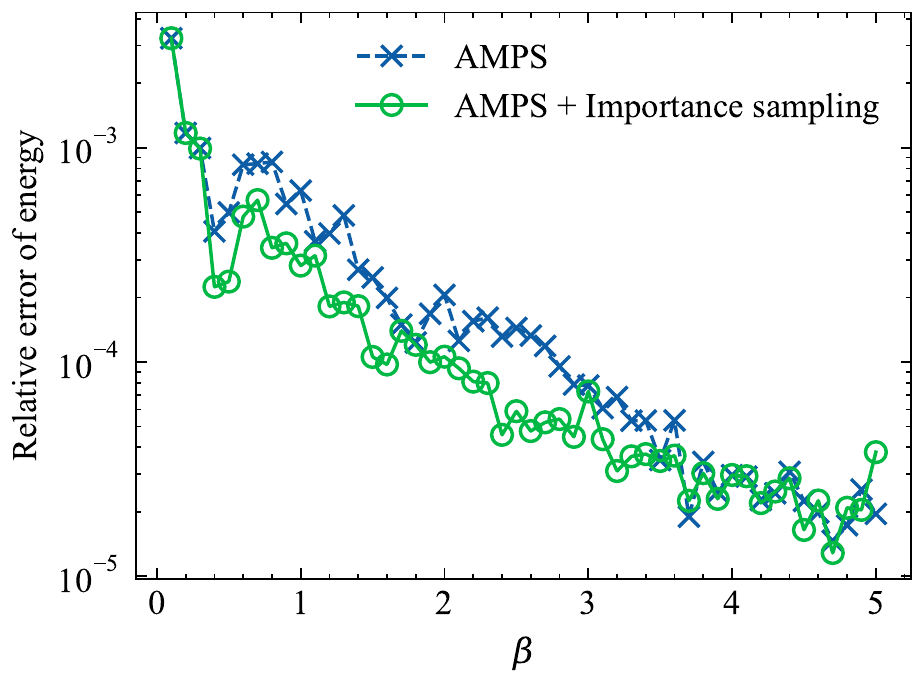}
    \caption{For the SK model with 20 spins, we show the relative error for estimating the energy $\langle E \rangle _p$ by (a) directly drawing samples from the AMPS and (b) applying neural importance sampling.}
    \label{fig:importance_sampling}
\end{figure}

The samples $\{\s\}$ drawn from $q(\s;\btheta)$ are unbiased to the AMPS distribution itself, but biased to the SK model, i.e. the Boltzmann distribution $P(\s)=\exp{(-\beta E(\s))}/Z$. However, this problem can be resolved by reweighting the samples. It is also called neural importance sampling in~\cite{nicoli2020asymptotically}.
In more detail, the expectation of an observable $\mathcal{O}(\s)$ can be computed as
\begin{align}
\langle \mathcal{O}(\s)\rangle_p &= \sum_\s p(\s)\mathcal{O}(\s)\nonumber\\
&=\sum_\s q(\s;\btheta)\frac{p(\s)}{q(\s;\btheta)}\mathcal{O}(\s)\nonumber\\
&=\left. \mathbb E_{\s\sim q}\left[\frac{e^{-\beta E(\s)}}{q(\s;\btheta)}\mathcal{O}(\s)\right]\right / \mathbb E_{\s\sim q}\left[\frac{e^{-\beta E(\s)}}{q(\s;\btheta)}\right]
\end{align}
In Fig.~\ref{fig:importance_sampling}, we show that neural importance sampling has alleviated the bias and can give a more accurate estimation of the energy $\langle E(\s) \rangle _p$.

\subsection{Comparing with direct tensor network contraction method}
\begin{figure}[!htbp]
    \centering
    \includegraphics[width=0.8\linewidth]{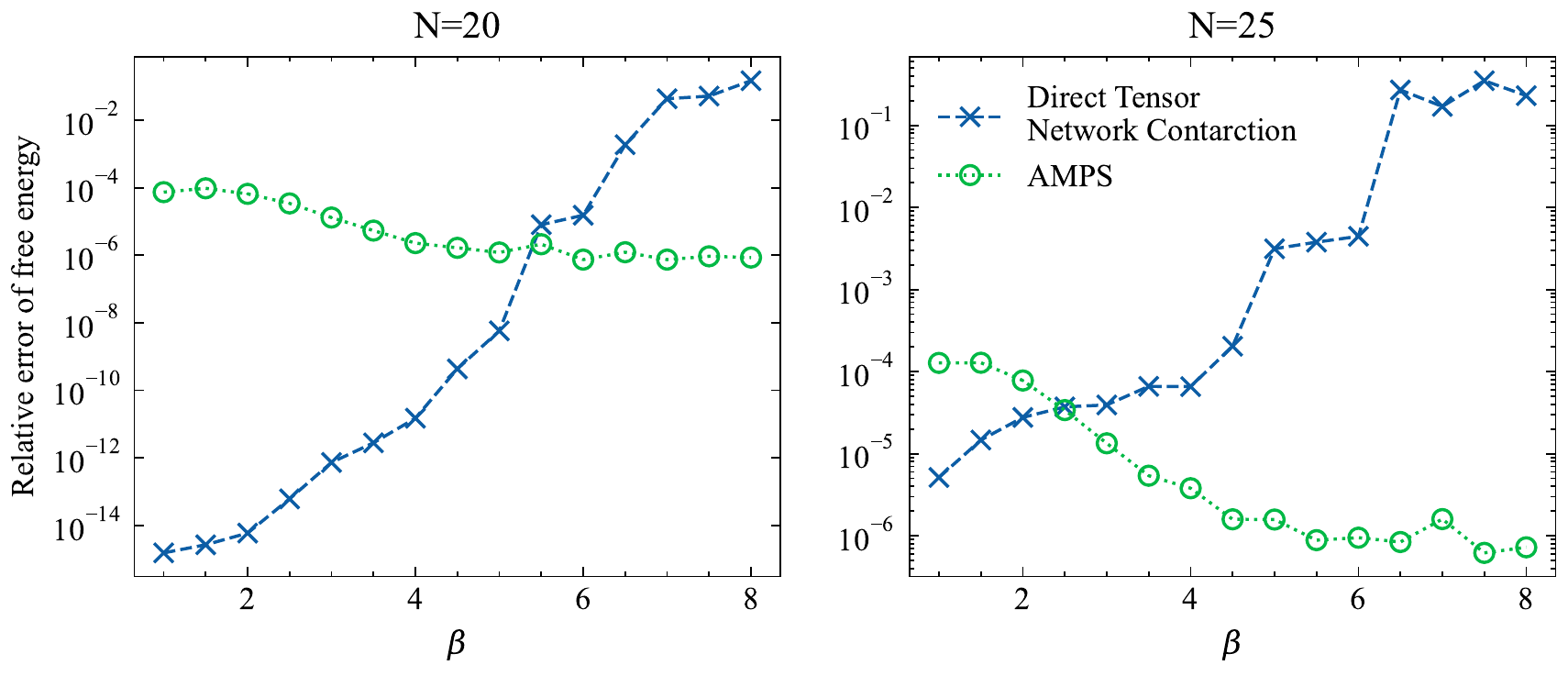}
    \caption{For the SK model with 20 and 25 spins, we compare the relative error of free energy for AMPS with direct tensor network contraction~\cite{pan2020contracting}.}
    \label{fig:compare-catn}
\end{figure}

The evaluation of the partition function $Z$ (free energy $F=-1/\beta \ln Z$) can also be translated into contracting the corresponding tensor network:
\begin{equation}
    Z = \sum_{\s}e^{-\beta E(\s)} = \sum_{\s} \prod_{\langle ij \rangle} e^{\beta s_i s_j}.
\end{equation}
We have compared the relative error of free energy made by both AMPS and the direct tensor network contraction method~\cite{pan2020contracting} in Fig.~\ref{fig:compare-catn}. We can see that with 20 spins, AMPS significantly outperforms the direct tensor network contraction method at low temperatures. This is because direct tensor network contraction needs to compress an exponential large tensor during the contraction process, which is possible at high temperatures but very difficult at low temperatures.

\subsection{Performance on larger systems}
\begin{figure}[!htbp]
    \centering
    \includegraphics[width=0.8\linewidth]{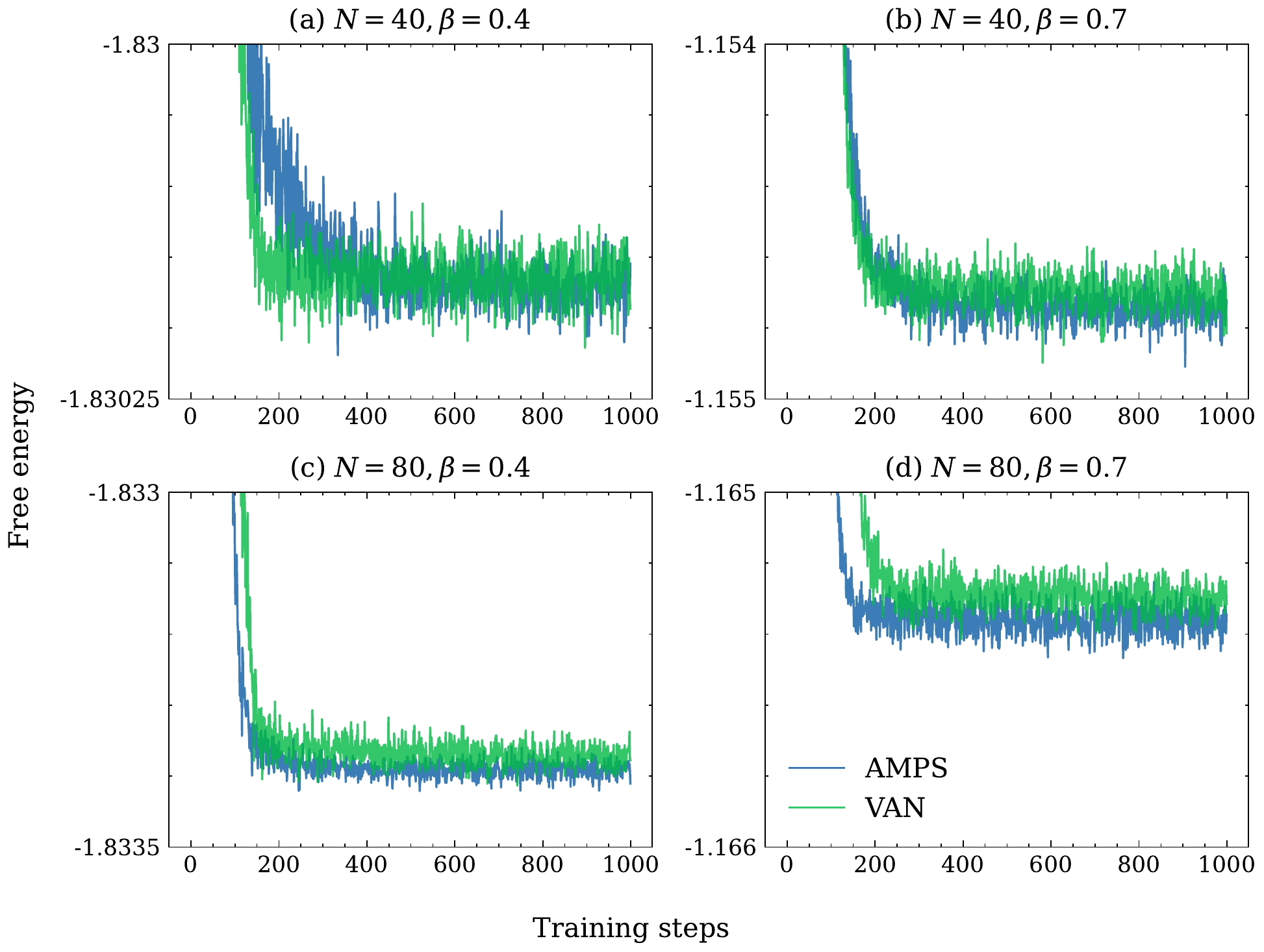}
    \caption{Free energy during the training process of AMPS and VAN on a SK model with $N=40/80$, $\beta=0.4/0.7$. The bond dimension of AMPS model is 4 and the depth and width of VAN are 2 and 10 respectively.}
    \label{fig:sk-scaling}
\end{figure}

In Fig.~\ref{fig:sk-scaling}, we plot the free energy during the training process of AMPS and VAN on a SK model with $N=40/80$, $\beta=0.4/0.7$. We can see from the figure that AMPS model has exhibited faster convergence and obtained lower free energy compared to the VAN. A more detailed and comprehensive study on larger systems ($N>100$) requires more training steps, searching for the optimal hyper-parameters as well as more GPU hours, which we leave for future work.

\end{document}